\documentclass[10pt, conference, compsocconf]{IEEEtran}

\usepackage{graphicx}
\usepackage{epstopdf}
\usepackage{float}
\usepackage{subfig}
\usepackage{cite}
\usepackage{paralist}
\usepackage{algorithmic}
\usepackage{amsmath}
\usepackage[linesnumbered,ruled]{algorithm2e}
\usepackage{xcolor}
\usepackage{diagbox}
\usepackage{booktabs}
\usepackage{multirow}
\usepackage{siunitx}

\begin{document}

\title{HPDedup: A Hybrid Prioritized Data Deduplication Mechanism\\ for Primary Storage in the Cloud}

\author{
\IEEEauthorblockN{Huijun Wu\IEEEauthorrefmark{1}\IEEEauthorrefmark{2}, Chen Wang\IEEEauthorrefmark{1}, Yinjin Fu\IEEEauthorrefmark{3}, Sherif Sakr\IEEEauthorrefmark{1}\IEEEauthorrefmark{2},  Liming Zhu\IEEEauthorrefmark{1}\IEEEauthorrefmark{2}, Kai Lu\IEEEauthorrefmark{4}}
\IEEEauthorblockA{\IEEEauthorrefmark{1}Data61, CSIRO}
\IEEEauthorblockA{\IEEEauthorrefmark{2}The University of New South Wales, Australia}
\IEEEauthorblockA{\IEEEauthorrefmark{3}PLA University of Science and Technology, China}
\IEEEauthorblockA{\IEEEauthorrefmark{4}Science and Technology on Parallel and
Distributed Laboratory,\\ 
State Key Laboratory of High Performance Computing, \\
State Key Laboratory of High-end Server \& Storage Technology, \\
College of Computer, National University of Defense Technology, Changsha, China}
}

\maketitle

\begin{abstract}

Eliminating duplicate data in primary storage of clouds increases the cost-efficiency of cloud service providers as well as reduces the cost of users for using cloud services.
Most existing primary deduplication techniques either use inline caching to exploit locality in primary workloads or use post-processing deduplication running in system idle time to avoid the negative impact on I/O performance. However, neither of them works well in the cloud servers running multiple services or applications for the following two reasons: Firstly, the temporal locality of duplicate data writes may not exist in some primary storage workloads thus inline caching often fails to achieve good deduplication ratio. Secondly, the post-processing deduplication allows duplicate data to be written into disks, therefore does not provide the benefit of I/O deduplication and requires high peak storage capacity.
This paper presents HPDedup, a \textbf{\underline{H}}ybrid \textbf {\underline{P}}rioritized data \textbf{\underline{Dedup}}lication mechanism to deal with the storage system shared by applications running in co-located virtual machines or containers by fusing an inline and a post-processing process for exact deduplication. In the inline deduplication phase, HPDedup gives a fingerprint caching mechanism that estimates the temporal locality of duplicates in data streams from different VMs or applications and prioritizes the cache allocation for these streams based on the estimation. HPDedup also allows different deduplication threshold for streams based on their spatial locality to reduce the disk fragmentation. The post-processing phase removes duplicates whose fingerprints are not able to be cached due to weak temporal locality from disks. The hybrid deduplication mechanism significantly reduces the amount of redundant data written to the storage system while maintaining inline data writing performance. Our experimental results show that HPDedup clearly outperforms the state-of-the-art primary storage deduplication techniques in terms of inline cache efficiency and primary deduplication efficiency.
\end{abstract}


\IEEEpeerreviewmaketitle

\section{Introduction}
Data deduplication is a technique that splits data into small chunks and uses the hash fingerprints of these data chunks to identify and eliminate duplicate chunks in order to save storage space. Deduplication techniques have achieved great successes in backup storage systems~\cite{zhu2008avoiding}. However, significant challenges remain to apply deduplication techniques in primary storage systems mainly due to the low latency requirement in primary storage applications~\cite{el2012primary}. Recent studies show that duplicate data widely exists in the primary workloads ~\cite{el2012primary}~\cite{meyer2012study}~\cite{srinivasan2012idedup}. In the cloud computing scenario, the primary workloads of the applications running on the same machine are observed having high duplicate ratio as well~\cite{mao2014pod}. For a cloud datacentre, there are significant incentives to remove duplicates in its primary storage for cost-effectiveness and competitiveness.

The existing data deduplication methods for primary storage can be classified into two main categories based on when the deduplication is performed: \emph{inline} deduplication techniques \cite{wildani2013hands}\cite{srinivasan2012idedup}\cite{mao2014pod} and \emph{post-processing} deduplication techniques \cite{el2012primary}\cite{an2013offline}\cite{constantinescu2011mixing}. The former performs data deduplication on the write path of I/O requests to immediately identify and eliminate data redundancy, while the latter removes duplicate data in background to avoid the performance impact on the I/O. However, challenges remain for both of these two methods.

For inline deduplication, fingerprint lookup is the main performance bottleneck due to that the size of a fingerprint table often exceeds the size of the memory. While a backup storage system may be able to tolerate the delay of disk based fingerprint lookup, the deduplication system of a primary storage system has to rely on caching to satisfy the latency requirement of applications. The state-of-art techniques for inline primary deduplication  \cite{wildani2013hands}\cite{srinivasan2012idedup}\cite{mao2014pod} exploit temporal locality of primary workloads by maintaining an in-memory fingerprint cache to perform deduplication. These deduplication mechanisms do not ensure that all duplicate chunks are eliminated. We call them \emph{non-exact deduplication}. However, the temporal locality in primary workloads does not always exist \cite{tarasov2014dmdedup}\cite{yupdfs}. For the workloads with weak temporal locality, caching the unnecessary fingerprints not only wastes the valuable cache space but also compromises other workloads with good locality. iDedup~\cite{srinivasan2012idedup} also exploits spatial locality to alleviate data fragmentation on disks by only eliminating duplicate block sequences longer than a fixed threshold. However, when data streams from different sources have different spatial locality, a fixed threshold may fail to achieve good deduplication ratio or read performance. For primary storage in clouds, the differences of locality become a severe problem. Firstly, the weak temporal locality becomes more apparent in the cloud when multiple applications running in virtualized containers sharing the same physical primary storage system. Since deploying deduplication in each virtual machine often fails to detect the duplicates among different virtual machines, deduplication should be deployed at the host physical machine. The data streams from co-located VMs or applications may interfere with each other and destroy the temporal locality. It has significant impact on the fingerprint cache efficiency managed by existing caching policies. The stream interference problem has been addressed in backup deduplication by resorting the streams \cite{kaisersorted}. However, it is not applicable for primary storage systems because we cannot change the order of requests of primary workloads. Secondly, as shown in our experiments on real-world traces (in Section \ref{sec:arch}), different workloads show quite different spatial locality. Therefore, a fixed global threshold is not optimal for alleviating the disk fragmentation in primary storage in the clouds.

For post-processing deduplication techniques \cite{el2012primary}\cite{an2013offline}\cite{constantinescu2011mixing}. There are two main drawbacks: Firstly, duplicate chunks are written to disks before being eliminated. This makes deduplication not effective in reducing peak storage use. For SSD based primary storage in cloud architecture like hyper-converged infrastructure, this affects the lifetime of SSD devices.
Secondly, the competition between post-processing deduplication process and foreground applications on using resources such as CPU, RAM and I/O can be a problem when a large amount of duplicates has to be eliminated.

To avoid the limitations and exploit the advantages of inline and post-processing deduplication, in this paper, we fuse the two phases together and propose a hybrid data deduplication mechanism to particularly deal with deduplication in virtualized systems running multiple services or applications from different cloud tenants.
The goal is to achieve a good balance between I/O efficiency and storage capacity saving in primary storage deduplication. In the inline deduplication phase, we differentiate the temporal locality of different data streams using a histogram estimation based method. The estimation method periodically assesses the temporal locality of the data streams from different services/applications. Based on the estimation, we propose a cache replacement algorithm to prioritize fingerprint cache allocation to favor data streams with good temporal locality. The mechanism significantly improves cache efficiency in inline deduplication and reduces the workload in the post-processing deduplication phase. Moreover, we adjust the threshold for different data streams dynamically to alleviate the disk fragmentation while achieving high inline deduplication ratio. The post-processing deduplication phase only deals with relatively small amount of duplicate data blocks that are missed in cache in the inline deduplication phase. Compared to systems that purely rely on post-processing deduplication, a highly efficient inline deduplication process greatly reduces the storage capacity requirement and contention in system resources. 

Overall, this paper makes the following main contributions:

\begin{enumerate}
\item We propose a novel hybrid deduplication mechanism that fuses inline deduplication and post-processing deduplication together for primary storage systems shared by multiple applications/services. The mechanism is able to provide exact deduplication in comparison to many inline deduplication mechanisms while avoiding drawbacks of purely post-processing deduplication mechanisms.

\item We give a locality estimation based cache replacement algorithm that significantly improves the fingerprint cache efficiency in primary storage deduplication systems. The estimation method is able to exploit locality in individual data streams for cache hit rate improvement.

\item We evaluate our mechanism using traces generated from real-world applications. The result shows that the proposed mechanism outperforms the state-of-art inline and post-processing deduplication techniques. e.g., HPDedup improves the inline deduplication ratio by up to 39.70\% compared with iDedup in our experiments. It also reduces up to 45.08\% disk capacity requirement compared with the state-of-art post-processing deduplication mechanism in our evaluation.
\end{enumerate}

The remaining of this paper is organized as follows: Section~\ref{sec:background} describes the background information and motivations for our approach; Section~\ref{sec:arch} presents the design of HPDedup;  Section~\ref{sec:cache} introduces how to differentiate the locality of data streams in deduplication; Section~\ref{sec:evaluation} presents the detailed results of our experimental evaluation; Section~\ref{sec:related} reviews related work and Section~\ref{sec:concl} concludes the paper.

\section{Background and Motivation}\label{sec:background}
In this section, we present the background and key observations that motivate this work.
\begin{figure*}[thb]
\centering
\subfloat[\textbf{\emph{FIU-mail trace}}]{\includegraphics[width=2.33in]{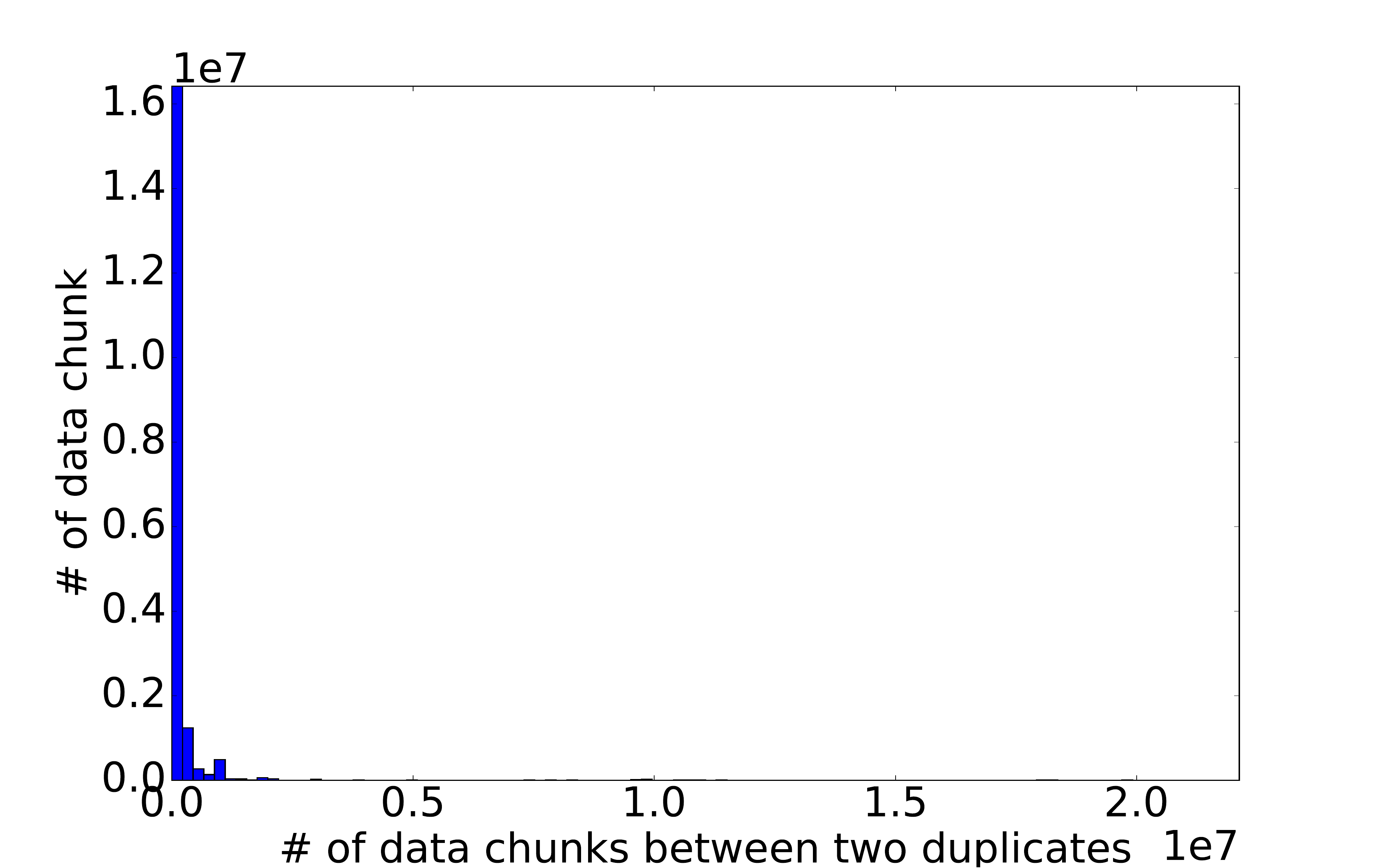}%
\label{fiu_mail}}
\hfill
\subfloat[\textbf{\emph{FIU-web trace}}]{\includegraphics[width=2.33in]{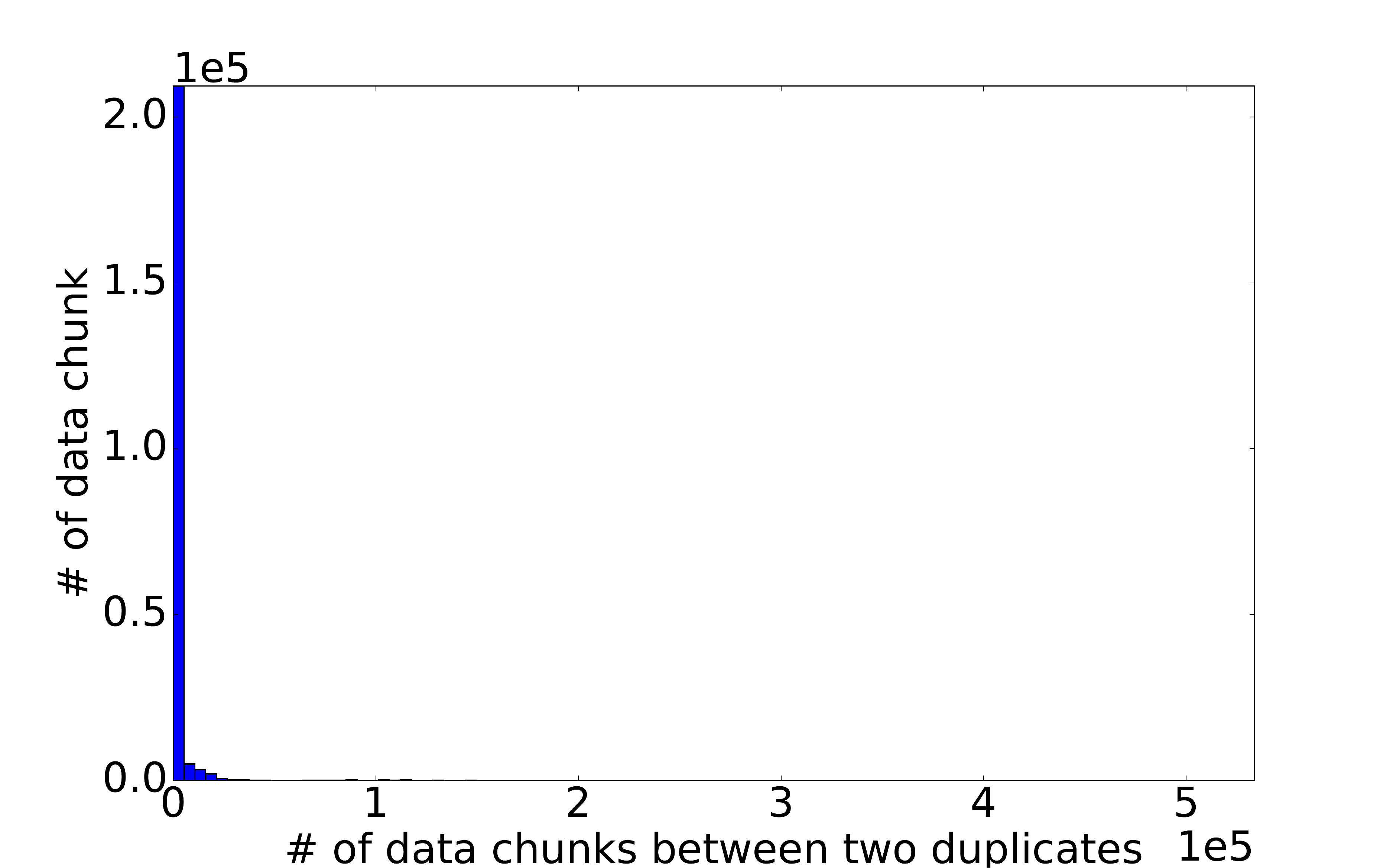}%
\label{fiu_webmail}}
\hfill
\subfloat[\textbf{\emph{Cloud-FTP trace.}}]{\includegraphics[width=2.33in]{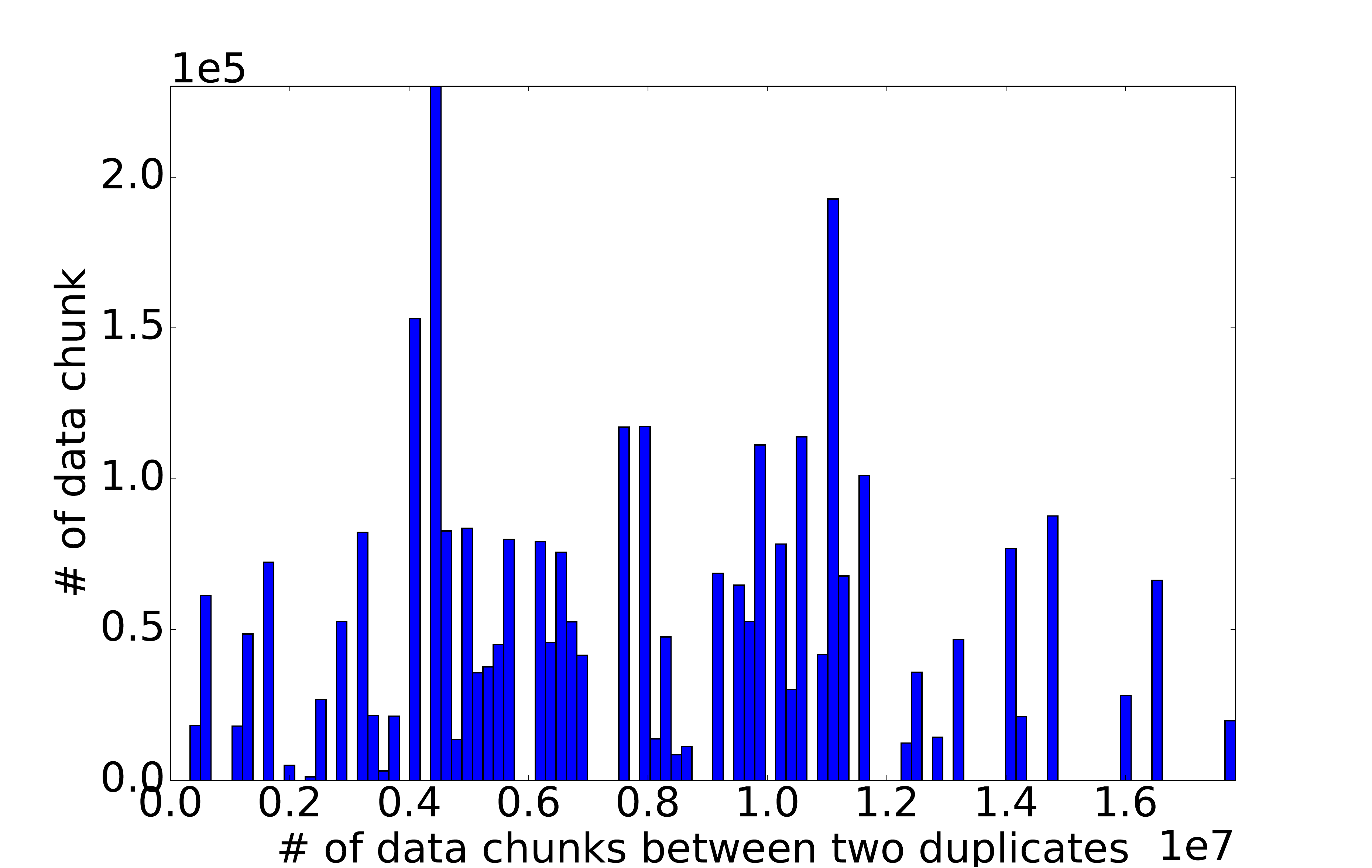}%
\label{our_file_server}}
\caption{Temporal locality analysis for three I/O traces. The x-axis is the number of data blocks between two adjacent occurrences of the same data block. i.e., for a I/O sequence "abac", each letter represents a data block. The number of data blocks between two adjacent occurrence of "a" is 1. }
\label{locality_exists}
\end{figure*}

\begin{figure*}[thb]
\centering
\subfloat[\textbf{\emph{LRU}}]{\includegraphics[width=2.3in]{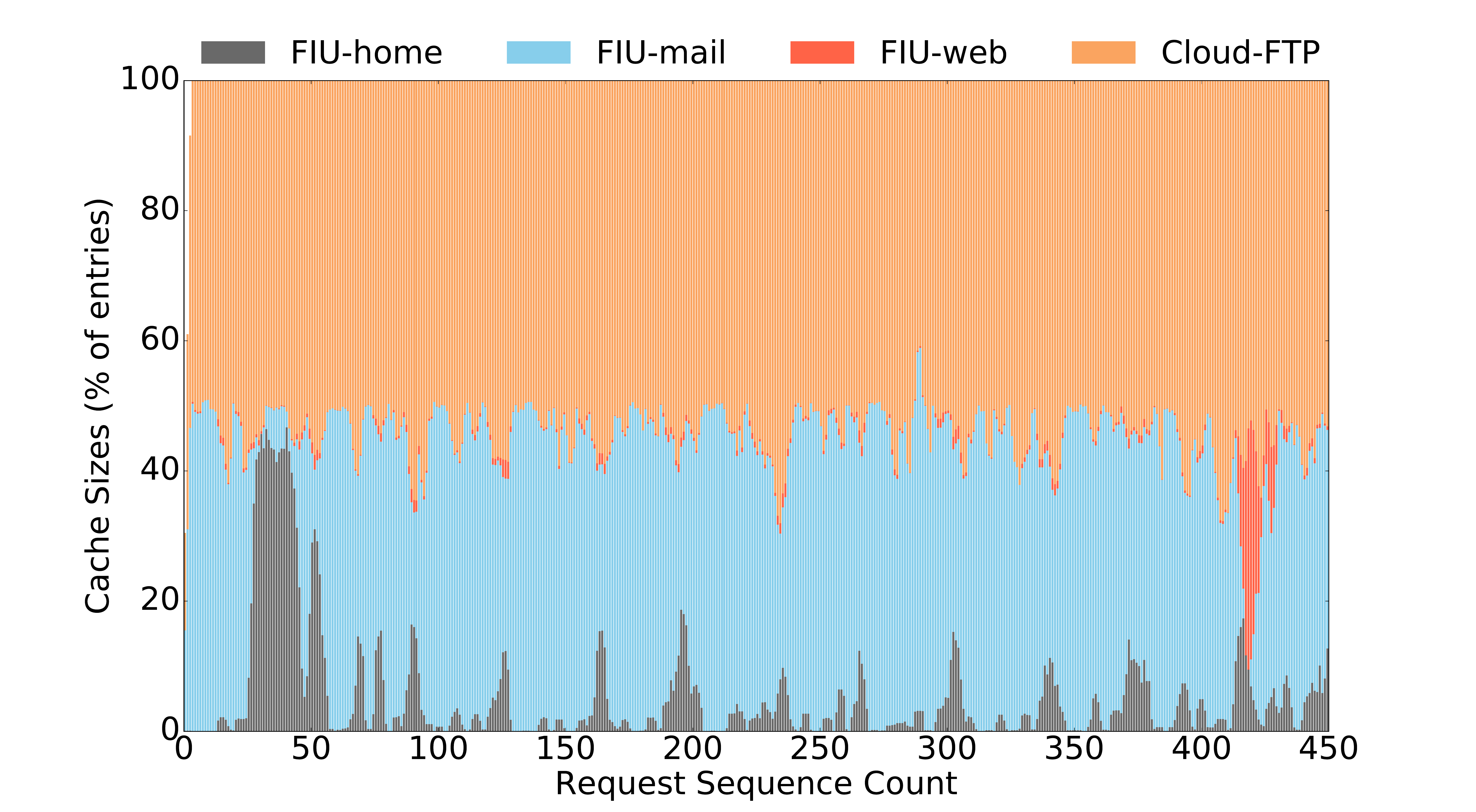}%
\label{lru_sample}}
\hfill
\subfloat[\textbf{\emph{LFU}}]{\includegraphics[width=2.3in]{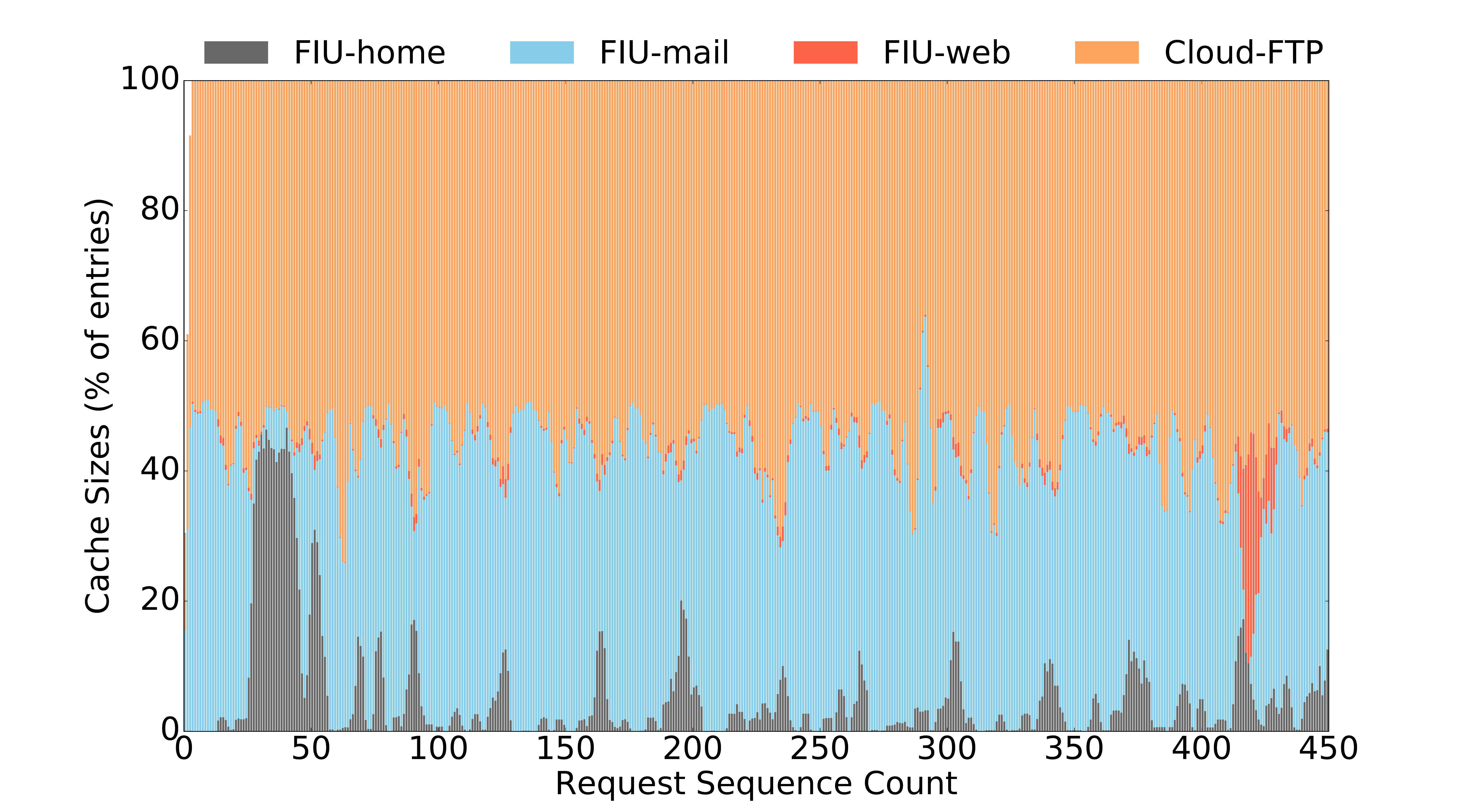}%
\label{lfu_sample}}
\hfill
\subfloat[\textbf{\emph{ARC}}]{\includegraphics[width=2.3in]{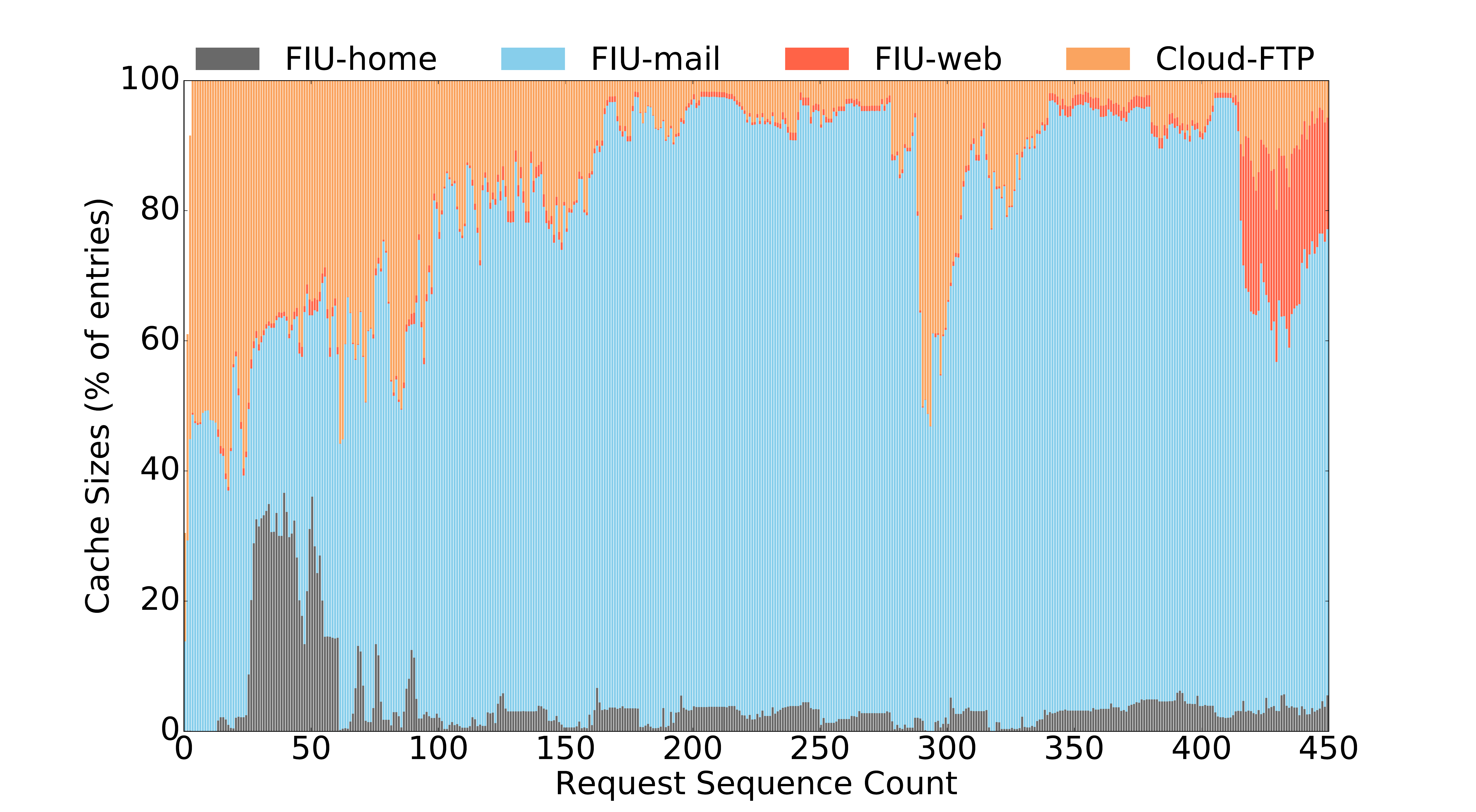}%
\label{arc_sample}}
\caption{The size of cache entries occupied by each data stream. The x-axis is the request sequence count while the y-axis is the percentage of cache occupied by different streams. The area of different colors indicates the cache resources used by each data stream.}
\label{fig:cache_size_sample}
\end{figure*}

\subsection{Deduplication for Primary Storage in Clouds}\label{subsec:dedupe_in_clouds}
Virtualization enables a cloud provider to accommodate applications from multiple users to run on a single physical machine while providing isolation between these applications. Recently, container techniques like Docker \cite{merkel2014docker} further reduce the overhead of using virtual machines to isolate user applications, thus support running more applications simultaneously on a physical machine.

Although providing much benefits in improving the resource sharing efficiency, increasing number of applications from different users sharing the same machine raises challenges to primary storage deduplication. In typical configurations, the cloud software stack such as OpenStack \cite{rhoton2014openstack} map the data volumes of VMs to persistent block storage devices connected by networks. It is impractical to achieve deduplication within a container or a virtual machine due to the overhead of storage device access. Moreover, it would fail to identify the duplicates among different VMs or containers. It is only feasible to detect duplication in the host's hypervisor that manages I/O requests from VMs or containers. As the file information in VMs or containers is not available in the underlying hypervisor, we design our deduplication mechanism to be based on the block level in the hypervisor. A similar architecture has also been used in an existing post-processing primary deduplication technique \cite{paulo2016efficient}.

\subsection{\textit{Temporal locality} Affects Efficiency of Fingerprint Cache}\label{subsec:temporal_locality_motivation}

Existing primary storage deduplication techniques often exploit temporal locality through fingerprint cache in an attempt to detect most of duplicates in the cache. However, some recent studies reveal that the locality may be weak in the primary storage workloads~\cite{yu2016pdfs}.

We evaluate the temporal locality with real-world traces, which contain a 24-hour I/O trace from a file server running in the cloud as well as the FIU trace~\cite{koller2010deduplication} commonly used in deduplication research. The file server is used for data sharing among a research group consisting of 20 people. We denote the file server trace as Cloud-FTP, and FIU mail server trace as \emph{FIU-mail} and FIU Web server trace as \emph{FIU-web}.

As shown in Figure \ref{locality_exists}, the average distance between two adjacent occurrences of a data block in both \emph{FIU-mail} and \emph{FIU-web} trace is small and highly skewed, indicating good locality. For the Cloud-FTP trace, the temporal locality is weak. The temporal locality of duplicates in primary storage systems varies among different applications. 

We further evaluate the cache efficiency for the three different workloads when they arrive at a storage system within the same time frame. The cache replacement algorithms we use in our evaluation include LRU (Least Recently Used), LFU (Least Frequently Used) and ARC (Adaptive Replacement Cache). The three cache replacement policies exploit the recency, frequency and the combination of both of workloads, respectively.

\begin{table}[h]
\centering
\caption{Workload Statistics of the 2-hour traces.}
\label{2hour_workload_stats}
\begin{tabular}{@{}cccc@{}}
\toprule
Trace     & Request number & Write request ratio & Duplicate writes \\ \midrule
Cloud-FTP & 2293424         & 84.15\%             & 387140                 \\
FIU-mail  & 1961588         & 98.58\%             & 1633424                \\
FIU-web   & 116940          & 49.36\%             & 30534                  \\
FIU-home  & 293605           & 91.03\%               & 32688                   \\ \bottomrule
\end{tabular}
\end{table}

We extract two-hour traces from the three FIU traces (10am-12am on the first day.) and the Cloud-FTP trace we collect. The characteristics of the two-hour traces are shown in Table \ref{2hour_workload_stats}. We mix these traces according to the timestamps of requests to simulate an I/O pattern of multiple applications on a cloud server. We set the cache size to 32K entries. Figure \ref{fig:cache_size_sample} illustrates the actual percentage of cache occupied by each data stream.
\begin{table}[h]
\centering
\caption{Duplicates detected under different cache replacement algorithms.}
\label{duplicates_found}
\begin{tabular}{@{}lllll@{}}
\toprule
Cache Policy & FIU-home & FIU-mail & FIU-web & Cloud-FTP \\ \midrule
LRU          & 20568     & 399622   & 16667   & 11977     \\
LFU          & 19984     & 381157   & 16072   & 11072     \\
ARC          & 22248     & 1119355  & 12245   & 467       \\ \bottomrule
\end{tabular}
\end{table}

 Table \ref{duplicates_found} shows the number of duplicate blocks detected by each cache policy. Under the LRU and LFU cache replacement algorithm, the cache allocated to Cloud-FTP stream is above 2/3 of the maximum cache capacity, but the number of duplicates detected in the stream is less than 2\% of the overall duplicates. Under the ARC cache replacement algorithm, the Cloud-FTP stream is allocated less portion of cache, however, only 467 duplicates are detected. This experiment shows that data streams with weak temporal locality of duplicates result in poor cache efficiency and fail to detect most of duplicates in the inline deduplication process.

When duplicates cannot be effectively detected through cache lookup, they are written to disks, which results in extra storage space requirement in capacity planning. It is therefore important to improve cache efficiency when locality of duplicates is not guaranteed.

\section{The Design of HPDedup}\label{sec:arch}
The inline or post-processing deduplication alone is difficult to satisfy the deduplication ratio and latency requirement of a primary storage system. However, the two techniques complement each other. Fusing them together to form a hybrid deduplication system is able to achieve \emph{exact deduplication} with satisfactory write performance. Particularly, the caching in inline deduplication not only speeds up fingerprint lookup, but also reduces the amount of data written to disks and relieves the burden of handling large amount of duplicates in the post-processing phase. On the other hand, the post-processing deduplication is able to detect duplicates missed out in the inline cache, therefore achieves \emph{exact deduplication}. Including a post-processing phase potentially relaxes the inline cache size requirement as well.

In another word, a hybrid deduplication system is able to achieve a balance between the I/O performance and deduplication ratio, which is essential for inline deduplication of primary storage. This motivates the architecture design of HPDedup.
In the following, we first give the hybrid architecture, and then describe the inline phase and post-processing phase of HPDedup.

\subsection{HPDedup Architecture Overview}\label{subsec:hpdedup_arch_paragraph}
\begin{figure}[htb]
\centering
\includegraphics[width=3.20in]{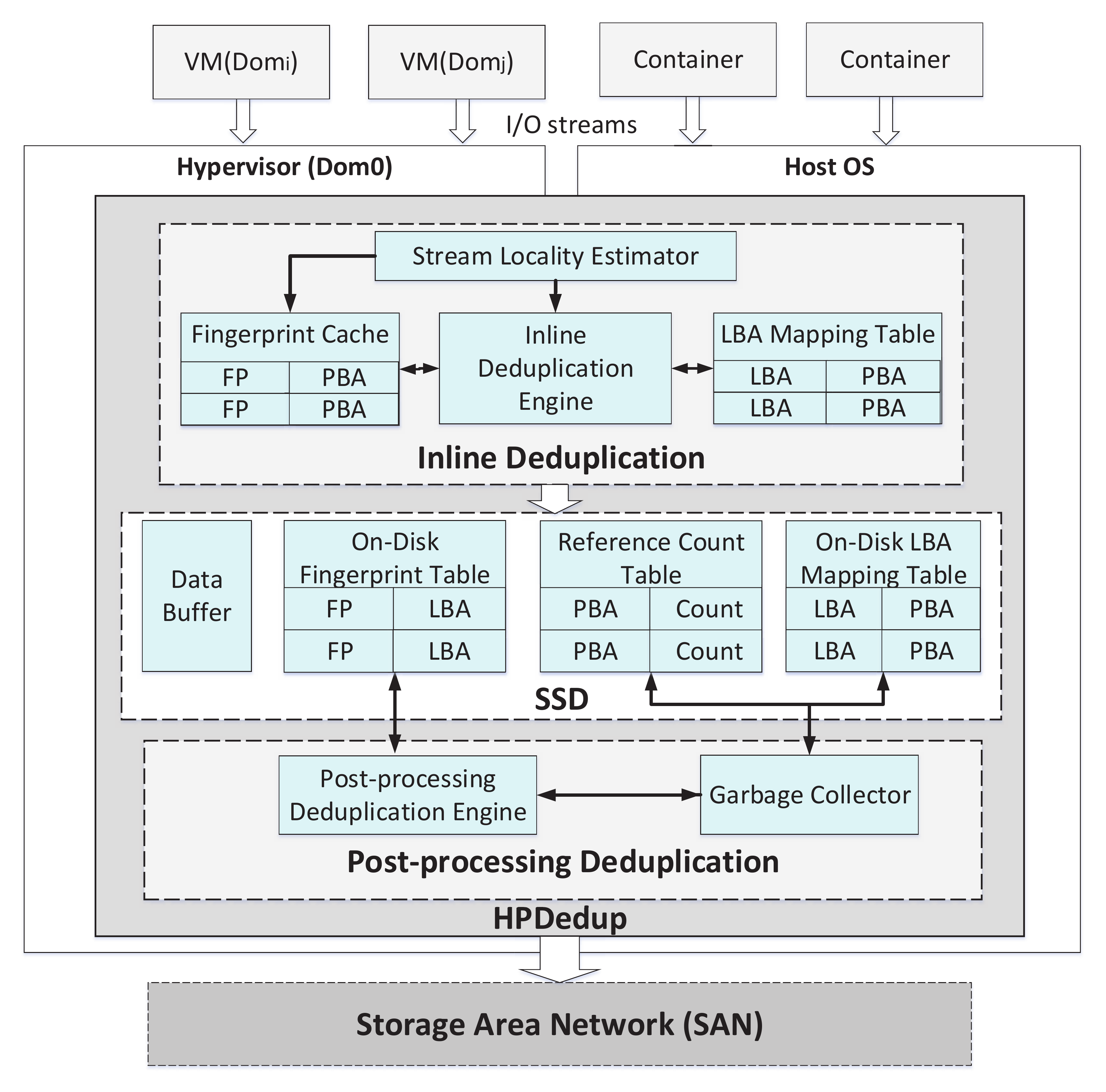}
\caption{System architecture of HPDedup.}
\label{HPDedup_arch}
\end{figure}
Virtualization is a core technology that enables the cloud computing. Running multiple virtual machines in a physical machine is a common practice in cloud datacentres. In cloud software stack such as OpenStack~\cite{rhoton2014openstack},  the storage volumes of virtual machines are often mapped to persistent block storage devices. It is impractical to implement deduplication inside a virtual machine while data streams from co-located VMs are written to the same physical device. The main reason is that performing deduplication in each VM is not able to remove the duplicates across VMs. We therefore place our deduplication mechanism at the block device level.

For the cloud scenario, the deduplication mechanism can be implemented inside the hypervisor that manages the I/O requests from VMs running on top of it. Some existing post-processing primary deduplication techniques~\cite{paulo2016efficient} place their deduplication mechanisms at the same level.

Figure~\ref{HPDedup_arch} illustrates the system architecture of HPDedup. A number of storage devices are connected to the server via SAN or through similar storage environments. The hypervisor (e.g., Xen) is responsible for translating LBA (logical block address) to PBA (physical block address) for block I/O requests from VMs running on top of it. HPDedup works at the hypervisor level to eliminate duplicate data blocks. For multiple containers running on the same host, HPDedup can be deployed at the block device level of the host machine. For simplicity, we will mainly use the hypervisor setting to describe the design of HPDedup in the rest of this paper.

\subsection{Inline Deduplication Phase}
In the inline deduplication phase, HPDedup maintains an \emph{in-memory fingerprint cache} that stores the fingerprint and PBA mapping to avoid slow disk-based fingerprint table search, and an \emph{LBA mapping table} that stores the mapping between LBAs and PBAs of blocks. The LBA mapping table is stored in NVRAM to avoid the data loss. The inline deduplication of data streams is performed in the \emph{inline deduplication engine}: the fingerprint of each data block is computed by a cryptographic hash function, like MD5 or SHA-1. The deduplication engine then looks up the block fingerprint in the fingerprint cache. The \emph{stream locality estimator} is responsible for monitoring and estimating both the temporal and spatial locality for the data streams coming from different VMs or containers. The temporal locality estimation is used for optimizing the hit rate of the fingerprint cache while the spatial locality estimation adjusts the deduplication threshold for data streams to reduce the disk fragmentation.

When the fingerprint of the incoming data block is found in the fingerprint cache, an entry of the LBA of the coming block and the corresponding PBA will be created and added into the LBA mapping table if such an entry does not exist, otherwise nothing is done because it is a duplicate write. If the block fingerprint is not found in the fingerprint cache, the data block is written to the underlying primary SAN storage. In this process, the data is staged in the data buffer in SSD for performance consideration. We use D-LRU \cite{li2016cachededup} algorithm to manage the data buffer in SSD to store recently accessed data by exploiting temporal locality.

When a data block is written to the underlying primary storage, the corresponding metadata associated with this data block including its fingerprint, LBA and PBA mapping as well as the reference count is updated in three tables in the SSD: on-disk fingerprint table, on-disk LBA mapping table and reference count table. The duplicates whose fingerprints are not cached in fingerprint cache will be eliminated in the post-processing phase.

\subsection{Post-Processing Deduplication Phase}\label{subsec:combining}
In the post-processing deduplication phase, the post-processing deduplication engine scans the on-disk fingerprint table and identifies duplicates. Note that duplicates identified in this phase are not in the fingerprint cache, thus they are not processed by inline deduplication phase. The entries with duplicate fingerprints are then removed while the corresponding LBAs are mapped to the same PBA in LBA mapping table. The reference count to the original PBAs containing the same data is decremented and the disk space is further claimed by the garbage collector. After post-processing, the unique data blocks in data buffer of SSD are organized into fixed-sized coarse-grained objects and flushed to the underlying persistent store.

\section{Differentiate Data Stream Locality in Deduplication}\label{sec:cache}
In this section we describe how to differentiate the temporal and spatial locality among different data streams to improve the efficiency of primary deduplication in the cloud. Both temporal and spatial locality estimation are performed in the stream locality estimator of the inline deduplication module. Specifically, the temporal locality of duplicates in a data stream is used to guide the allocation of fingerprint cache to the data stream in order to achieve higher inline deduplication ratio. The spatial locality of a stream is used to achieve a balance between the inline deduplication ratio and the read performance. We first describe how to measure and estimate the temporal locality of duplicates in data streams in \ref{subsec:temporal_locality}, and then describe how to manage the fingerprint cache based on the temporal locality measurement in \ref{subsec:cache_management}. Thirdly, we discuss how to handle the disk fragmentation based on the difference of spatial locality among data streams in \ref{subsec:spatial_locality}.

\subsection{Temporal Locality Estimation for data streams}\label{subsec:temporal_locality}

The \emph{temporal locality of duplicates} characterizes how soon duplicates of a data block may arrive in the system in a data stream. A good temporal locality indicates duplicate data blocks generally are close to each other while a weak locality indicates that duplicate blocks are often far away from each other or there are few duplicates in the data stream.

To measure the temporal locality of duplicates, we introduce a metric called \textit{Local Duplicate Set Size (LDSS)}. \textit{LDSS} of a stream is defined as the number of duplicate fingerprints in last $n$ contiguous data blocks arriving before a given time. Here, we call $n$ \textit{estimation interval}. 

To use \emph{LDSS} to guide fingerprint cache allocation, we need to predict the \textit{LDSS} of the future arrivals of data blocks of the data stream. A common approach is to use the historical \textit{LDSS} values to predict the future \emph{LDSS} of the data stream. To obtain a historical \textit{LDSS} value from a data stream, a naive way is to count all distinct fingerprints for each data stream and their occurrences within an \textit{estimation interval}. However, this incurs a high memory overhead which is close to the cache capacity because all the fingerprints need to be recorded. To address the problem, the \textit{stream locality estimator} uses the reservoir sampling algorithm \cite{vitter1985random} to sample fingerprints from a data stream, and then estimate \textit{LDSS} from these samples using the unseen estimation algorithm \cite{valiant2013estimating}.

Reservoir sampling algorithm assumes an unknown number of fingerprints in a data stream and guarantees that each fingerprint in the data stream has an equal chance to be sampled. In our implementation, each element in the \textit{sampling buffer} is a pair containing a fingerprint and its occurrence count.

The unseen estimation algorithm is able to estimate the unseen data distribution based on the histogram of the samples of observed data. For HPDedup, the unseen estimation algorithm is used to estimate \textit{LDSS} of data streams based the sampled fingerprints from these streams. We refer readers to \cite{valiant2013estimating} for more details about the theoretical aspect of estimation algorithms. Here, we give a high-level description of using the unseen estimation algorithm to estimate the \textit{LDSS} values for a data stream.

Consider the storage system handles \textit{M} data streams, denoted by $S_{1}$,$S_{2}$,...,$S_{M}$ from \textit{M} VMs and the \textit{estimation interval} size is \textit{n},
the goal of the temporal locality estimation is to collect $k$ fingerprint samples from the last $n$ write requests of each stream and compute the \textit{LDSS} values for these streams based on fingerprint samples. 

Specifically, after sampling, we denote the number of sampled fingerprints coming from stream \textit{i} by $N_{i}$. By using unseen estimation algorithm, we can accurately estimate the number of unique writes (denoted by $u_{i}$) in stream $S_{i}$ among last $n$ write requests in the mixed stream. Then the estimated \textit{LDSS} for stream $i$ can be denoted by $LDSS_{i}$ as below:
\[
LDSS_{i} = N_{i} - u_{i}
\]
whereas $N_{i}$ is the total number of write requests for stream $i$ in the \textit{estimation interval}.

The estimation of $u_{i}$ as well as the calculation of ${LDSS_{i}}$ is shown in Algorithm~\ref{unseen_alg}. Before discussing the algorithm, we introduce a concept named \textit{Fingerprint Frequency Histogram (FFH)}. A \textit{FFH} of a set of fingerprints $F$ is a histogram $f = \{f_{1},f_{2}...\}$ where $f_{j}$ is the number of distinct fingerprints that appear exactly $j$ times in $F$.

We derive the \textit{FFH} from the \textit{sampling buffer} to estimate the $LDSS_{i}$ of data stream $i$. We use $H_{s}$ to denote the \textit{FFH} of the samples and $H$ to denote the \textit{FFH} of the whole estimation interval for stream $i$. According to the unseen estimation algorithm, we then compute the transformation matrix $T$ by a combination of binomial probabilities about the chances an item is drawn a certain times. The expected histogram $H_{s}^{'}$ for sampled data blocks can be computed by $H_{s}^{'} = T \cdot H$. To solve the equation and get $H$, we minimize the distance between $H_{s}$ and $H_{s}^{'}$ which are the observed histogram and expected histogram of sampled data blocks, respectively. Once obtained  $H$, we are able to compute the $LDSS_{i}$  for the data stream.

\begin{algorithm}
    \SetKwInOut{Input}{Input}
    \SetKwInOut{Output}{Output}
    \Input{$H_{s}$ -- The \textit{FFH} for samples in stream $i$; \\ $N_{i}$ -- the number of write requests of\\ stream $i$ in the estimation interval.}
    \Output{Estimated $LDSS_{i}$ of data stream $i$}

	Compute matrix \textit{T} by binomial probabilities.\\
	$H_{s}^{'}$ for samples is computed by $H_{s}^{'} = T \cdot H$\\

	Linear programming: \\
	\quad Objective function:  $min(\Delta(H_{s}, H_{s}^{'}))$, in which
	\quad\quad $\Delta(H_{s}, H_{s}^{'}) = \Sigma_{i} \frac{1}{\sqrt{H_{s}[i] + 1}} |H_{s}[i] - (T \cdot H)[i]|$. \\
	\quad under constraints: \\
	\quad\quad $\Sigma_{i} H[i] = N$\\
	\quad\quad $\forall i \quad H[i] > 0$\\
	\quad return $LDSS_{i} = N_{i} - \Sigma_{i}H[i]$

    \caption{Temporal Locality Estimation Algorithm}
    \label{unseen_alg}
\end{algorithm}

For some streams which have few write requests during the \textit{estimation interval}, it is not necessary to run unseen estimation algorithm to estimate the \textit{LDSS}. The \textit{LDSS} of these streams are set to a small value for simplicity.

\subsection{LDSS Estimation Based Fingerprint Cache Management}\label{subsec:cache_management}
We use the \textit{LDSS} of different streams to guide the cache allocation for these streams. The fingerprints from a data stream with higher predicted \textit{LDSS} is more likely to be kept in the cache than those from a data stream with lower \textit{LDSS}. As mentioned earlier, we use the historical \textit{LDSS} values which are accurately estimated by the unseen algorithm to predict the \textit{LDSS} of streams. We use self-tuned double exponential smoothing method to predict the \textit{LDSS}  values. Using the estimated $LDSS(w-2),LDSS(w-1),... $, we can predict $LDSS(w)$ where $w$ is the next \textit{estimation interval}. The predicted \textit{LDSS} is used to guide the fingerprint cache management as follows.

Firstly, we propose a cache admission policy that the fingerprints from streams with very low \textit{LDSS} would not be cached if there exists streams with much higher \textit{LDSS}. This strategy can avoid caching the fingerprints of data streams containing compressed data or other forms of compact data. The cache of each stream can be managed by any cache replacement policies.

Secondly, for the fingerprints already cached,  we assign an evict priority value $p_i$ to data stream $i$, denoted by:
\[
p_i = \frac{1}{LDSS_{i}(w)}
\] The evict priorities are mapped to adjacent non-overlapping segments in a segment tree. Specifically, stream $i$ is represented by the segment [$\Sigma_{k=0}^{i-1}p_{k}$, $\Sigma_{k=0}^{i-1}p_{k} + p_{i}$).
When evicting a fingerprint from the cache, we generate a random number $r$ and find the segment $I$ to which $r$ belongs. We then evict one cache entry from the cache corresponding to interval $I$.

\begin{figure}[h]
\centering
\includegraphics[width=3.0in]{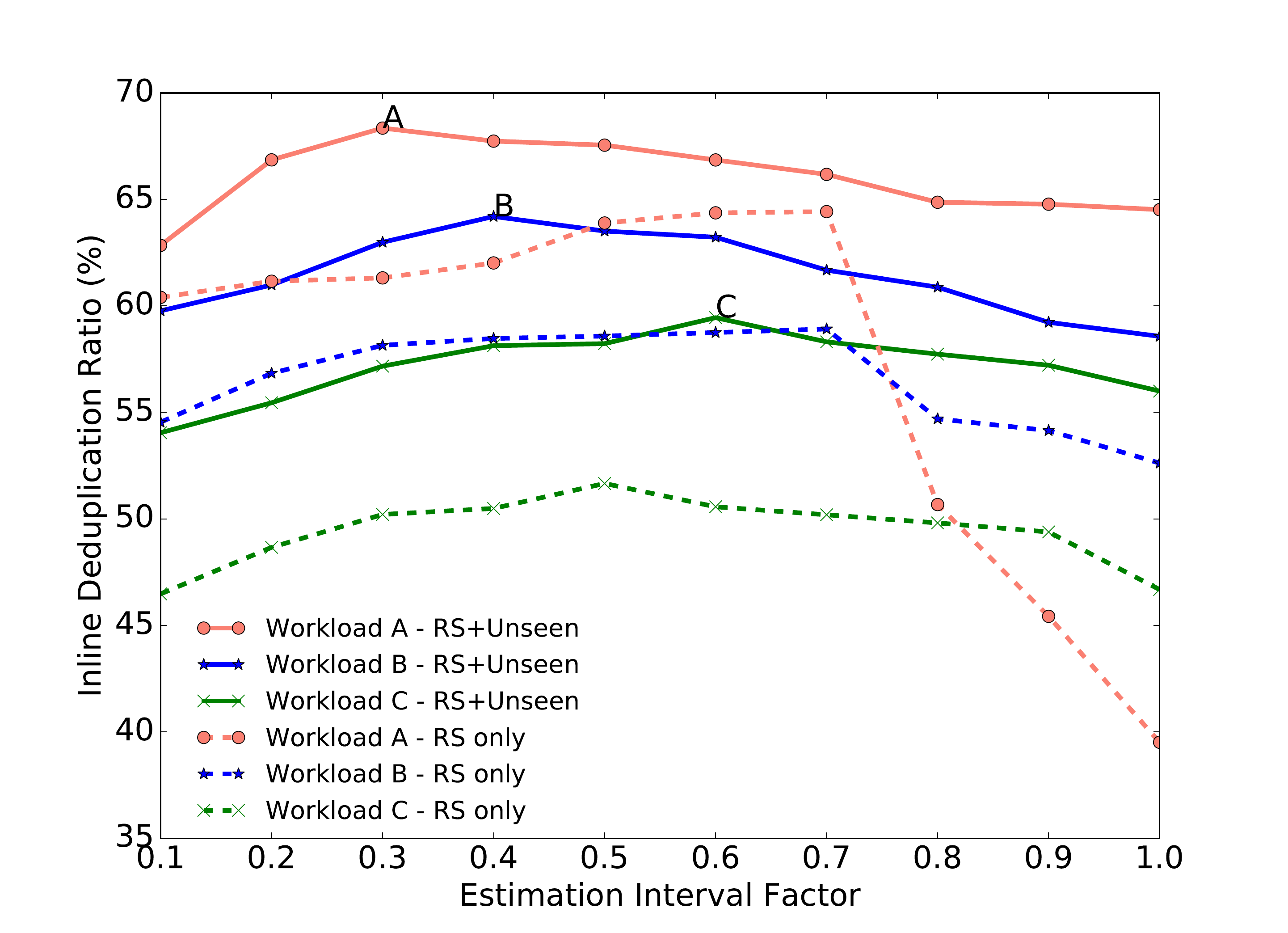}
\caption{Inline Deduplication ratio vs. Estimation Interval Factor for three different workloads.}
\label{fig:estimation_interval_factor}
\end{figure}

As the fingerprint cache management of HPDedup relies on the accurate \textit{LDSS} estimation of using unseen estimation algorithm, one may also think about directly estimating the \textit{LDSS} of data streams by the number of duplicate fingerprint samples sampled by reservoir sampling. 
Figure \ref{fig:estimation_interval_factor} compares the effectiveness of using RS-only (Reservoir sampling only, dash lines) or RS + Unseen (Reservoir sampling with unseen algorithm, solid lines) during the \textit{LDSS} estimation. It is clear that RS + Unseen based \textit{LDSS}  estimation is able to provide much higher inline deduplication ratio with a smaller estimation interval compared with RS-only method. This shows the effectiveness of temporal locality estimation using the unseen algorithm.

Moreover, a proper value of \textit{estimation interval} is important for achieving the good cache efficiency. Too large interval may include some out-dated information which cannot reflect the current temporal locality of duplicates for the workloads. Too small interval, on the other hand, cannot accurately capture the temporal locality. Since \textit{LDSS} is used to estimate the number of duplicates which can be detected in the fingerprint cache, the \textit{estimation interval} can be set to a factor of the number of fingerprint cache entries.  The solid lines in Figure \ref{fig:estimation_interval_factor} shows the inline deduplication ratio of HPDedup for three different workloads (details are shown in Section \ref{sec:evaluation}) while choosing different \textit{estimation interval factor}. The cache size is set to 160MB. From the workload A to C, the overall temporal locality for the workload decreases. The \textit{estimation interval factor} needs to be set to larger values for the workloads with worse temporal locality. Correspondingly, we can see that the optimal \textit{estimation interval factor} for workload A, B and C are 0.3, 0.4 and 0.6, respectively.  In practice, a good approximation is to set the \textit{estimation interval factor} to 1 - d where d is the historical inline deduplication ratio for the mixed streams.

The temporal locality estimation is triggered by the following three events: 1. the finish of an \textit{estimation interval}; 2. a significant drop of inline deduplication ratio; 3. the join or quit of virtual machines/applications.

\subsection{Spatial Locality Aware Threshold for Deduplication}\label{subsec:spatial_locality}
\begin{figure}[t]
\centering
\includegraphics[width=2.8in]{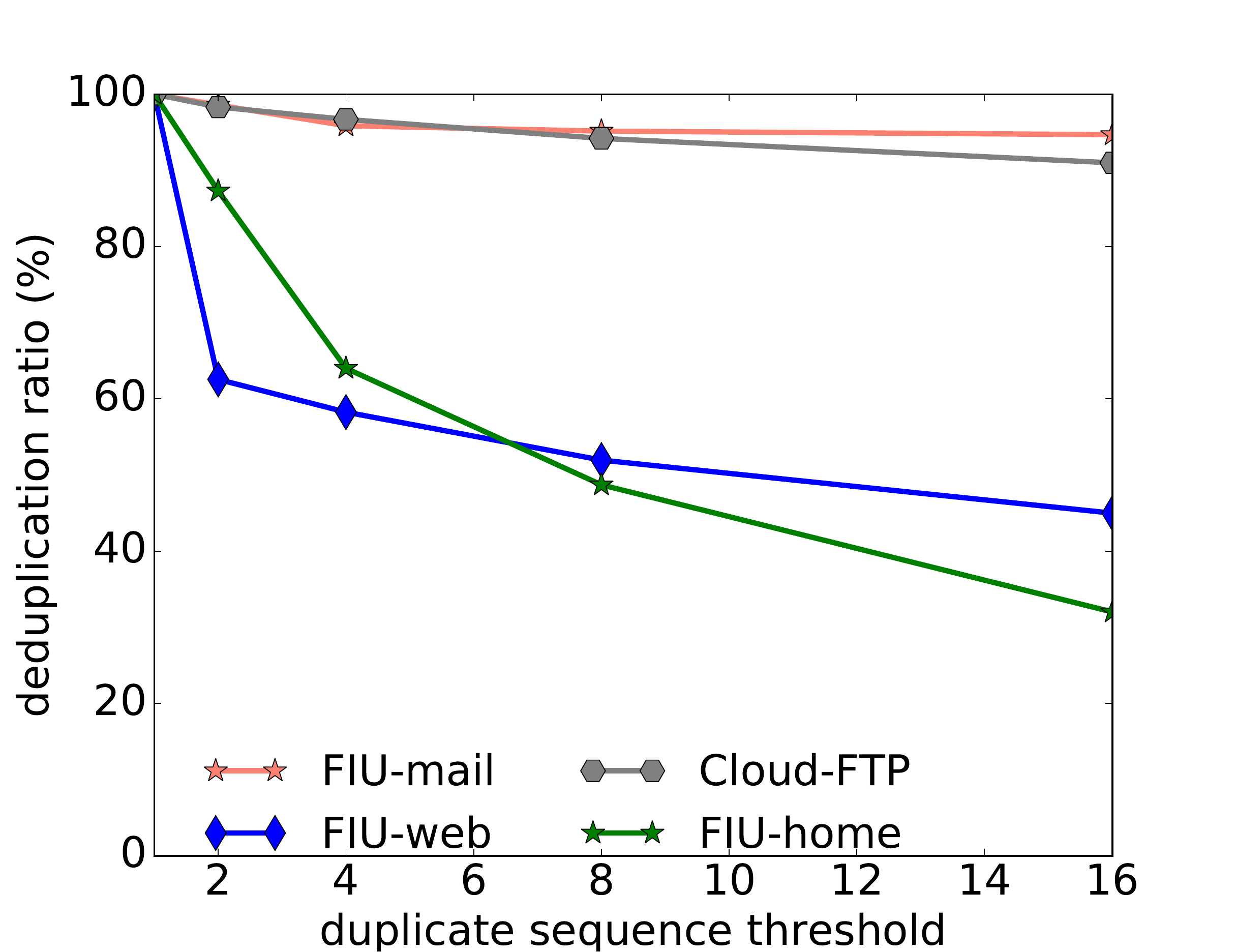}
\caption{Deduplication ratio vs. Threshold. Deduplication ratio versus threshold for different threshold of duplicate sequence length.}
\label{fig:dedupe_ratio_vs_threshold}
\end{figure}

To alleviate disk fragmentation problem of deduplication on data read, some primary storage deduplication techniques only eliminate duplicate block sequences with length greater than a given \textit{threshold}. However, as pointed by \cite{srinivasan2012idedup}, for the applications with many random I/Os, doing so may not find any duplicates. In the deduplication of primary storage systems in clouds, the spatial locality of data streams for different applications/services varies significantly. To explore the relationship between deduplication ratio and threshold, we analyze both the FIU traces and the trace we collect. As shown in Figure \ref{fig:dedupe_ratio_vs_threshold}, different workloads show different trends. When the threshold increases from 1 to 16, the inline deduplication ratio for FIU-mail and Cloud-FTP reduces by only 4.3\% and 9.1\%, respectively. The inline deduplication ratio for FIU-web drops by around 38.1\% when the threshold increases from 1 to 2. When the threshold is 16, the inline deduplication ratio is 43.1\% of that under a threshold of 1. For FIU-home trace, the inline deduplication ratio keeps dropping. When the threshold is set to 16, the inline deduplication ratio is only 32.0\% of that under a threshold of 1 .

Figure~\ref{fig:dedupe_ratio_vs_threshold} indicates that the threshold value should be adaptive to data stream characteristics. It is noteworthy that there exists a tradeoff between the write latency and read latency while choosing a proper threshold. Write operations prefer shorter threshold while read operations prefer longer threshold. For writes, long sequence indicates more comparisons before writing data blocks to disks. For reads, long threshold can avoid many random I/Os thus reducing the read latency.

HPDedup maintains two vectors $V_{w}$ and $V_{r}$ for each stream. $V_{w}$ is used to store the occurrence number for the largest length values of sequential duplicates. $V_{r}$ is used to store the occurrence number for the length values for sequential read. Both $V_{w}$ and $V_{r}$ have 64 items. For instance, if $V_{w}[3]=100$, there are 100 sequential duplicates with length 3 since $V_{w}$ is reset. If $V_{r}[3]=100$, there are 100 sequential reads with length 3 since $V_{r}$ is reset.

Initially, the threshold is 16. The two vectors collect data when requests come. When the threshold update is triggered, given the histogram vector $V_{w}$ and $V_{r}$, the threshold $T$ is computed by
\[
T = (1 - r) \cdot \overline{Len_{d}} + r \cdot \overline{Len_{r}}
\]
where $\overline{Len_{d}}$  and $\overline{Len_{r}}$ are the average length of duplicate block sequence and average read length, respectively. $T$, therefore, is the balance point of the read and write latency. $r$ is the read ratio among all requests. $\overline{Len_{d}}$ and $\overline{Len_{r}}$ are computed according to the data collected in $V_{w}$ and $V_{r}$, respectively. To cope with the changes of duplicate pattern for each stream, the two vectors is reset to all 0s when the total deduplication ratio decreases by over 50\% since the last threshold update.

\section{Evaluation}\label{sec:evaluation}

The prototype of HPDedup is implemented in C. To evaluate the performance of HPDedup, we use real-world traces to feed into HPDedup. We compare the performance of HPDedup with the following deduplication methods: locality based inline deduplication (iDedup \cite{srinivasan2012idedup}), post-processing deduplication schema (e.g.,\cite{el2012primary} \cite{paulo2016efficient}) and hybrid inline-offline deduplication schema DIODE \cite{tang2016diode}.

\subsection{Configuration}
The experiments are carried out in a workstation with Intel Core i7-4790 CPU , 32GB RAM and 128GB SSD + 1TB HDD. We use the \emph{FIU-home, FIU-web, FIU-mail}~\cite{koller2010deduplication} traces in our evaluation. These traces are from three different applications, namely remote desktop, web server and mail server respectively in FIU. Moreover, we also collect a trace from a cloud FTP server (Cloud-FTP) used by our research group. The trace is obtained by mounting a network block device (NBD) as the working device for the file server, from which we capture read/write requests through a customized NBD-server.

To the best of our knowledge, there is no available larger scale I/O traces containing both the fingerprint of data blocks and timestamps. We therefore use the four traces as templates to synthetically generate VM traces representing multiple VMs. Table \ref{workload_stats} shows the statistics of the four workloads. The arrival order of requests in these workloads are sorted and merged based on timestamps. The generated trace has the same I/O pattern with the original traces. For the traces generated from the same template, the content overlap is randomly set to 0\% - 40\% which is the typical data redundancies among users \cite{sunlong}.

\begin{table}[h]
\centering
\caption{Workload Statistics}
\label{workload_stats}
\begin{tabular}{@{}cccc@{}}
\toprule
Trace     & Num of requests & Write request ratio & Duplicate ratio \\ \midrule
Cloud-FTP & 21974156        & 83.94\%             & 20.77\%         \\
FIU-mail  & 22711277        & 91.42\%             & 90.98\%         \\
FIU-web   & 676138          & 73.27\%             & 54.98\%         \\
FIU-home  & 2020127         & 90.44\%            & 30.48\%         \\ \bottomrule
\end{tabular}
\end{table}

In our experiments, we simulate a cloud host running 32 virtual machines. As shown in Figure \ref{locality_exists}, FIU traces show better temporal locality compared with Cloud-FTP trace. We mix traces to form three workloads with different overall temporal locality.  Workload A contains 15 mail server traces, 5 FTP server traces and 8 remote desktop traces and 4 web server traces. Workload B contains 10 mail server traces, 10 FTP server traces, 6 remote desktop server traces and 6 web server traces. Workload C contains 5 mail server traces, 15 FTP server traces, 6 web server traces and 6 remote desktop server traces. The ratios of data size between the good-locality (L) traces and bad-locality (NL) traces are around 3:1, 1:1 and 1:3 for these three mixed workloads.

The \textit{estimation interval factor} is set to 0.5 at the beginning and is adjusted by the historical inline deduplication ratio dynamically for each workload.

\subsection{Cache Efficiency in Inline Deduplication}
\begin{figure*}[t]
\centering
\hfill
\subfloat[\textbf{\emph{Workload A (L:NL = 3:1)}}]{\includegraphics[width=0.333\textwidth]{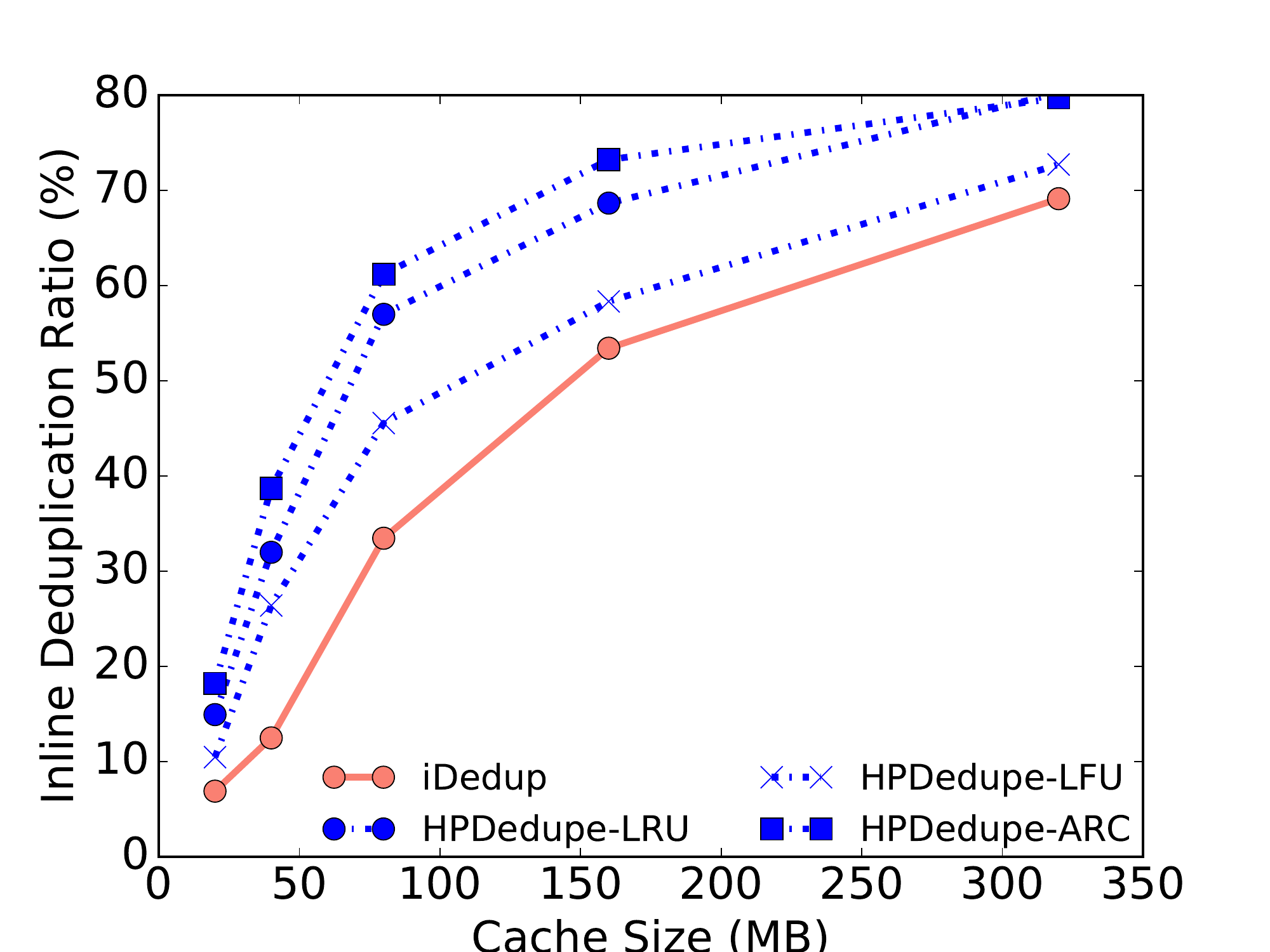}%
\label{3V1}}
\subfloat[\textbf{\emph{Workload B (L:NL = 2:2)}}]{\includegraphics[width=0.333\textwidth]{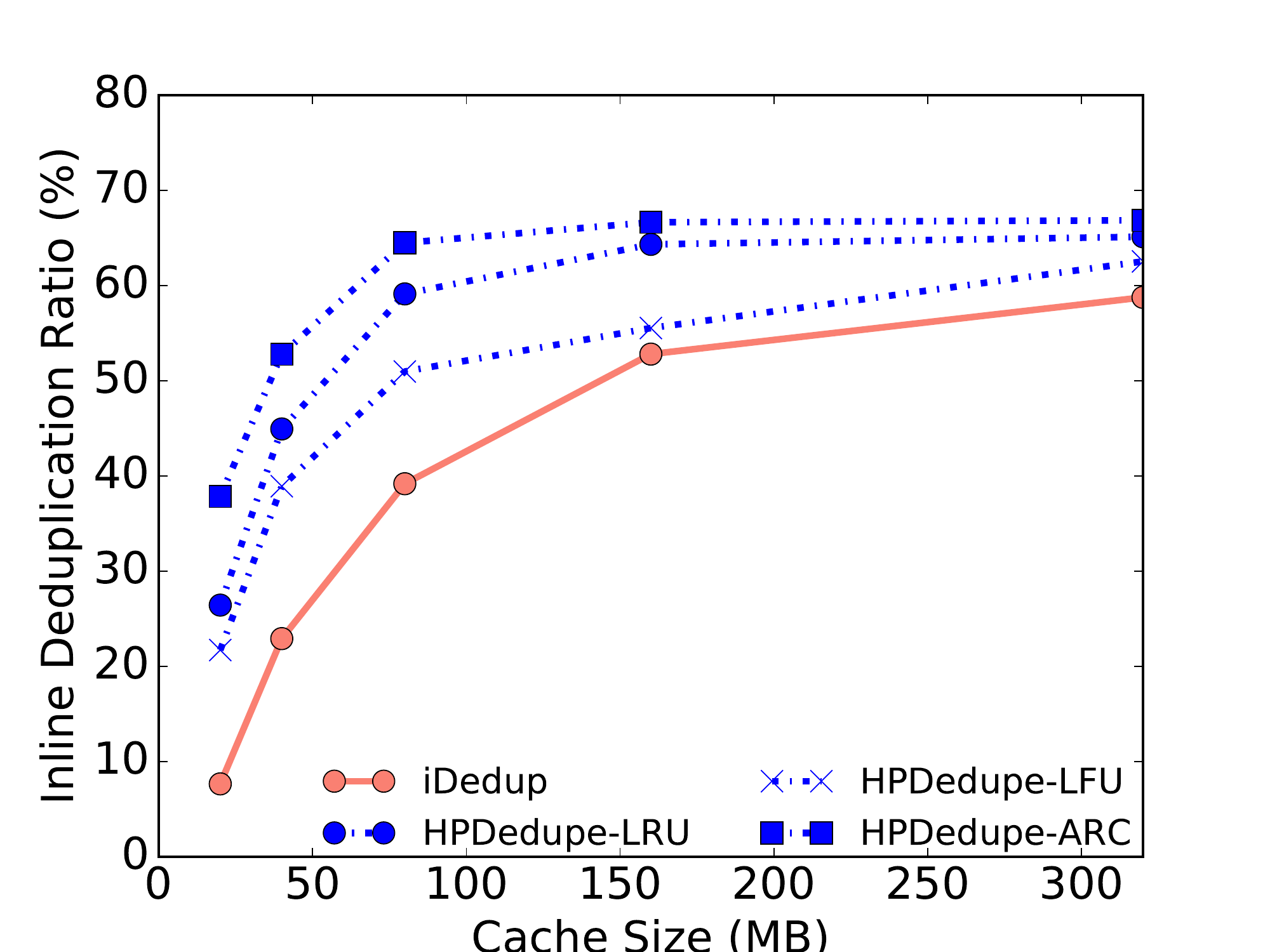}%
\label{2v2}}
\hfill
\subfloat[\textbf{\emph{Workload C (L:NL = 1:3)}}]{\includegraphics[width=0.333\textwidth]{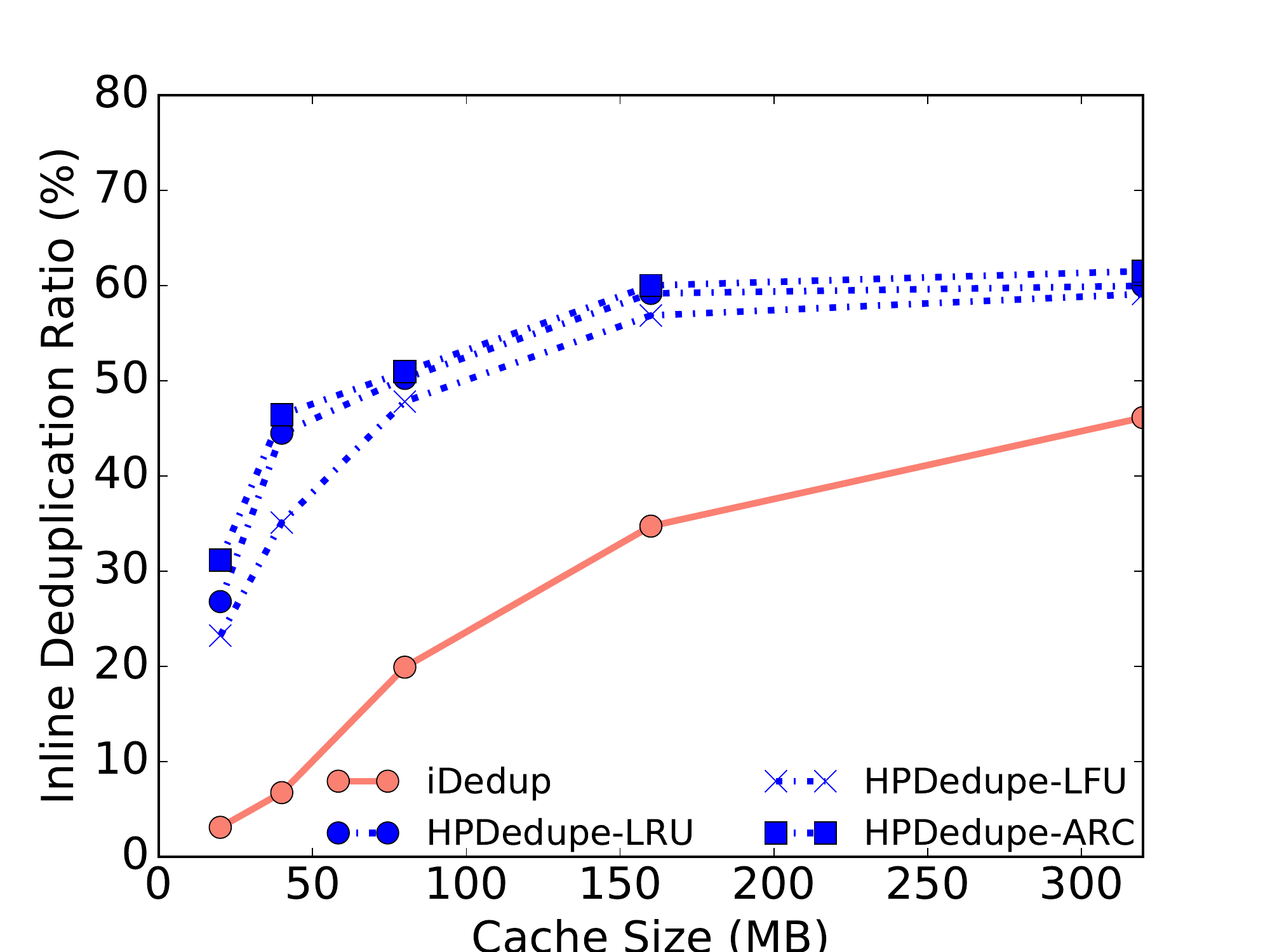}%
\label{1v3}}
\caption{Inline deduplication ratio vs. Cache size for iDedup and HPDedup with different cache replacement policies.}
\label{fig:inline_dedup_ratio}
\end{figure*}

We compare HPDedup with iDedup, a well-known inline deduplication system that makes use of temporal locality of primary workload. We replay the mixed workloads to simulate the scenario where multiple applications/services running on the same physical machine. Each I/O for the three workloads is a 4KB block and MD5 is used as the hash function to calculate the fingerprints.

Each entry of the deduplication metadata is about 64 bytes and contains the fingerprint and the block address. 
According to the size and footprint of the traces, the total memory size for fingerprint cache is set from 20MB to 320MB in the experiments.
The deduplication threshold is set to 4 for both iDedup and HPDedup.

Figure \ref{fig:inline_dedup_ratio} shows the inline deduplication ratio versus cache size for iDedup and HPDedup. Here, inline deduplication ratio is defined as the percentage of duplicate data blocks that can be identified by inline caching. For the cache replacement policy of each stream, LRU, LFU and ARC cache replacement policies are supported by HPDedup. LRU was claimed to be the best cache replacement policy by iDedup \cite{srinivasan2012idedup}.

When the portion of NL workload increases, the gap between iDedup and HPDedup becomes larger. HPDedup-LRU, HPDedup-LFU and HPDedup-ARC improve the inline deduplication ratio significantly compared with iDedup. For workload A, HPDedup-ARC improves the inline deduplication ratio by 10.58\% - 27.72\%. HPDedup-LRU improves the inline deduplication ratio of iDedup which also uses LRU cache by 8.04\% - 23.52\%. HPDedup-LFU shows less improvement (3.56\%-13.90\%) compared with HPDedup-ARC and HPDedup-LRU. Similarly, for workload B, HPDedup-ARC, HPDedup-LRU and HPDedup-LFU achieve 8.09\%-30.19\%, 6.36\%-22.02\% and 3.77\%- 16.02\% improvement of inline deduplication ratio, respectively. For workload C, HPDedup-ARC improves the inline deduplication ratio by 15.38\% - 39.70\%. HPDedup-LRU and HPDedup-LFU achieve 13.86\% - 37.75\% and 12.97\% - 28.37\% improvement, respectively.  The improvement is larger when the cache size is small due to cache resource contention. Moreover, HPDedup-ARC outperforms HPDedup-LRU and HPDedup-LFU because ARC cache replacement policy makes the size of T1 (LRU) cache and T2 (LFU) cache adaptive to the recency and frequency of the workloads. It is also noteworthy that with the increasing of non-locality workload share (from Workload A to C), the inline deduplication ratio improvement achieved by HPDedup becomes larger. The reason is that non-locality workloads provide more space for the optimization of HPDedup.

Note that for HPDedup-ARC, extra memory overhead is introduced by ARC caching replacement policy itself to track the evicted fingerprints and their metadata. The overhead is non-trivial for fingerprint cache. The analysis of overhead for obtaining statistics information about evicted fingerprints in existing cache replacement policies is out of the scope of this paper. We leave the discussion to the future work. In the following experiments, HPDedup-LRU is used by default. 

Overall, the weak locality in workloads results in low inline deduplication ratio. \textbf{With the locality estimation method in HPDedup, the allocation of inline fingerprint cache dynamically gives the streams with better locality higher priority. Hence, HPDedup improves the overall cache efficiency for multiple VMs/applications running on the same physical machine in the cloud.}

\subsection{Disk Capacity Requirement}
\begin{figure}[htb]
\vspace{-10pt}
\centering
\includegraphics[width=2.50in]{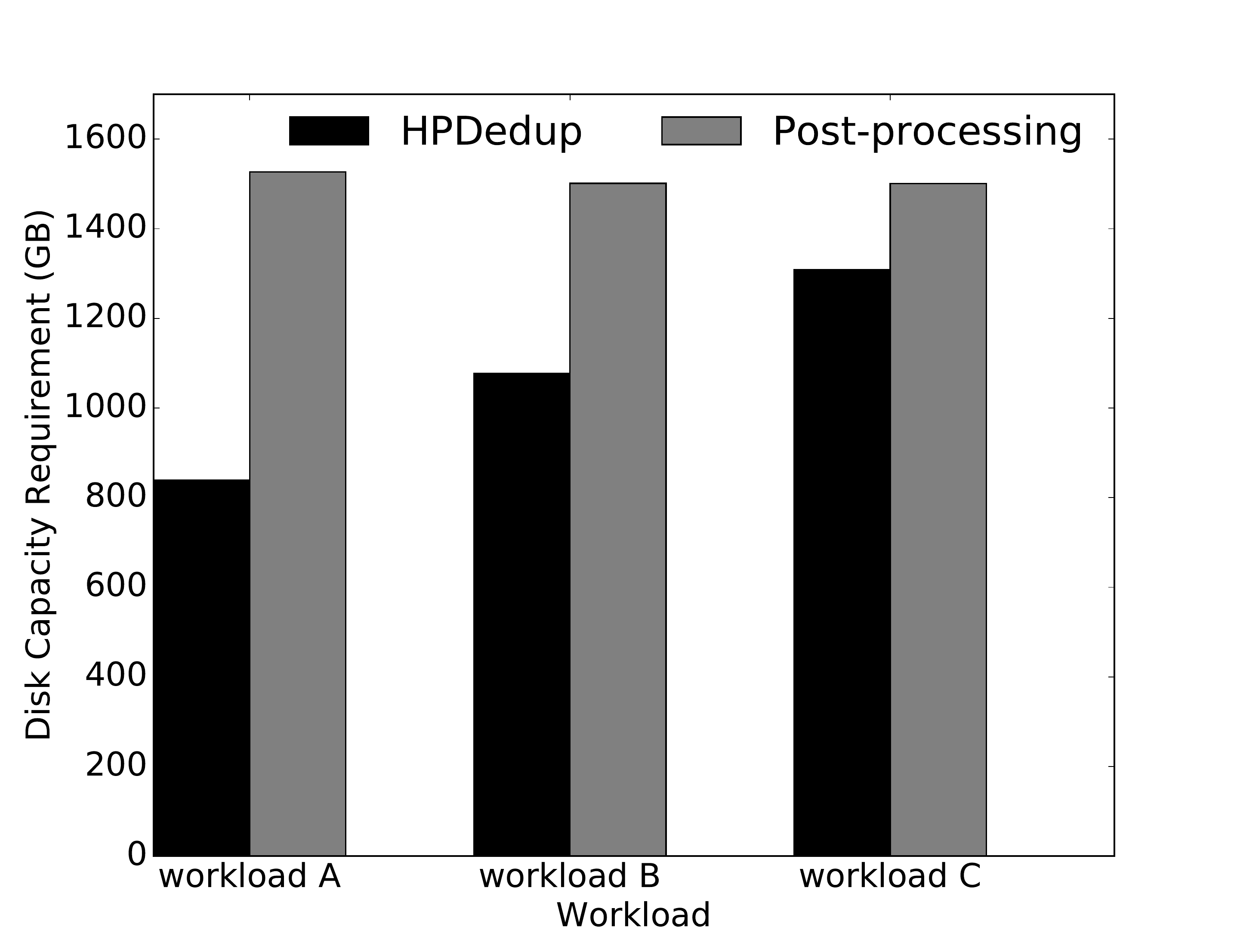}
\caption{The disk capacity requirements for HPDedup and post-processing deduplication schemas.}
\label{fig:offline_data_size}
\end{figure}
In this subsection, we compare the size of the data before performing post-processing deduplication for HPDedup and pure post-processing deduplication (e.g., \cite{el2012primary}). The size is also the maximum required disk size for the deduplication mechanisms. Figure \ref{fig:offline_data_size} shows the result of comparisons.

For HPDedup, the inline fingerprint size is set to 160MB and LRU cache replacement policy is used for simplicity. As shown in Figure~\ref{fig:offline_data_size}, HPDedup significantly reduces the disk capacity requirements for storage space. The data size has been reduced by 45.08\%, 28.29\% and 12.78\% for workload A, B and C, respectively. The better the locality is in the workload, the more duplicates can be detected in the inline phase and the more duplicate data writes can be eliminated. \textbf{This clearly shows the benefit of a hybrid deduplication architecture of HPDedup as hundred GBs of data writes can be reduced by only maintaining a 160MB inline fingerprint cache.}

\subsection{Average Hits of Cached Fingerprints}
Inline deduplication of HPDedup does not require disk access to identify duplicates therefore is faster than the post-processing based deduplication. We use \emph{average hits of cached fingerprints} as an indicator of inline deduplication performance. The indicator is obtained by monitoring the number of fingerprint entering the fingerprint cache. With a high average hits value, the inline deduplication is able to detect a large portion of duplicates and reduces the load of the more expensive post-processing deduplication.

In the following, we compare HPDedup with DIODE~\cite{tang2016diode} on this metric. DIODE uses file extensions to decide whether to perform inline deduplication on these files. Files are classified roughly into three different types. The inline deduplication process skips the type of files containing audio, video, encrypted and other compressed data (called P-Type in DIODE). We use a full inline deduplication method \cite{srinivasan2012idedup} as the baseline.

DIODE works at the file system level so that the information like file extensions are passed to the hypervisor layer in the form of hints, like the method used by \cite{mandal2016using}. The file type information of the Cloud-FTP trace is known so that the trace can be used to test DIODE. 
The files that are classified as P-Type are around 14.2\% of the whole trace in size. The FIU traces are classified into U-Type (unpredictable type), which will be processed by the inline deduplication, just like that in the evaluation setting of DIODE in \cite{tang2016diode}.  


\begin{table}[h]
\centering
\caption{The \emph{Average Hits of Cached Fingerprints} for baseline, DIODE and HPDedup.}
\label{fcu}
\begin{tabular}{@{}ccccc@{}}
\toprule
Schema                    & Cache Size & Workload A & Workload B & Workload C \\ \midrule
\multirow{2}{*}{Baseline} & 320MB      & 1.301      & 0.665      & 0.204      \\
                          & 160MB      & 0.781      & 0.551      & 0.148      \\ \midrule
\multirow{2}{*}{DIODE}    & 320MB      & 1.437      & 0.812      & 0.247      \\
                          & 160MB      & 0.805      & 0.598      & 0.181      \\ \midrule
\multirow{2}{*}{HPDedup}  & 320MB      & 1.812      & 4.024      & 5.918      \\
                          & 160MB      & 1.409      & 3.101      & 5.002      \\ \bottomrule
\end{tabular}
\end{table}

\begin{figure}[t]
\centering
\subfloat[\textbf{\emph{CentOS VM image}}]{\includegraphics[width=2.8in]{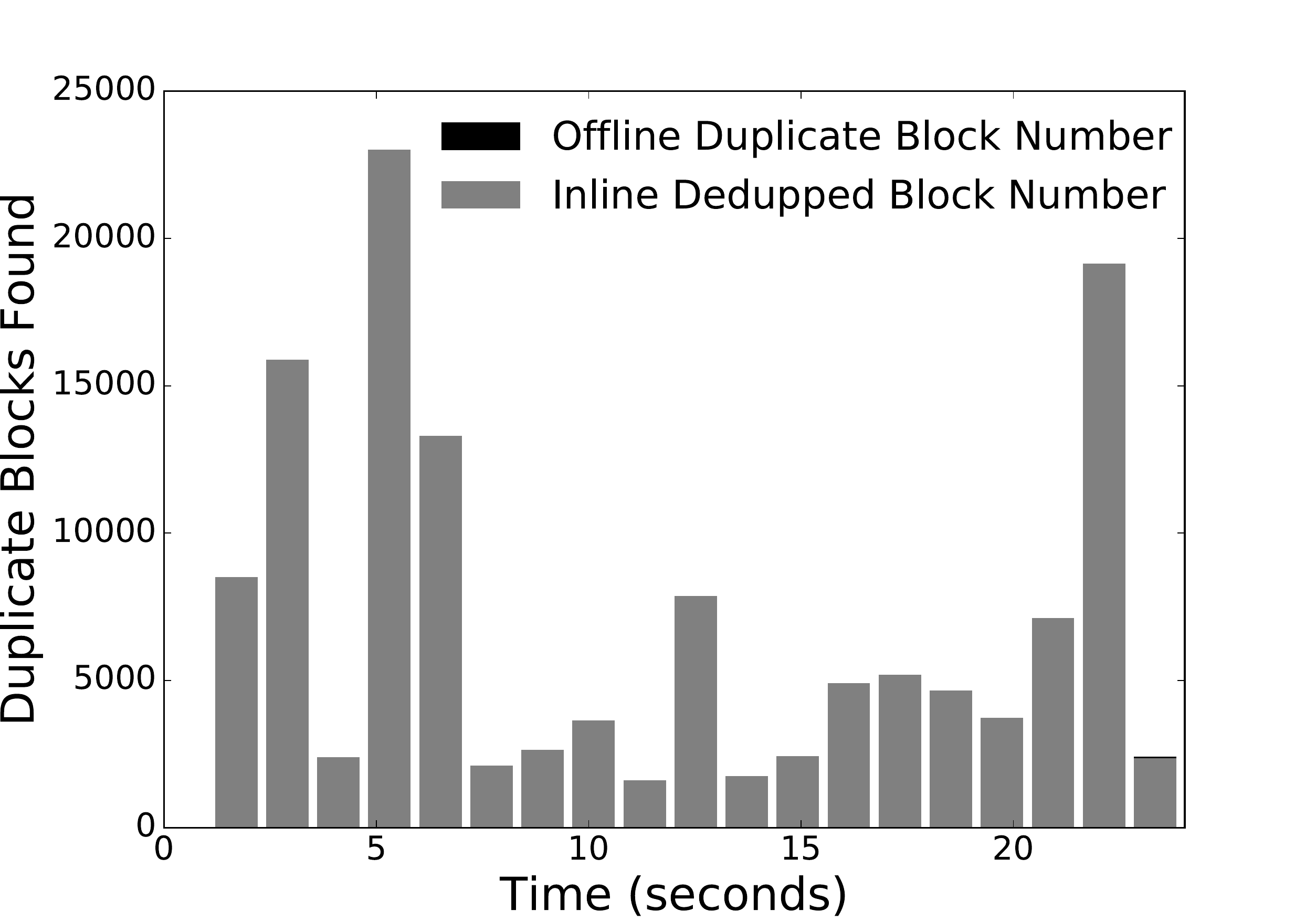}%
\label{centos}}
\vfill
\vspace{-10pt}
\subfloat[\textbf{\emph{Linux Source Code}}]{\includegraphics[width=2.8in]{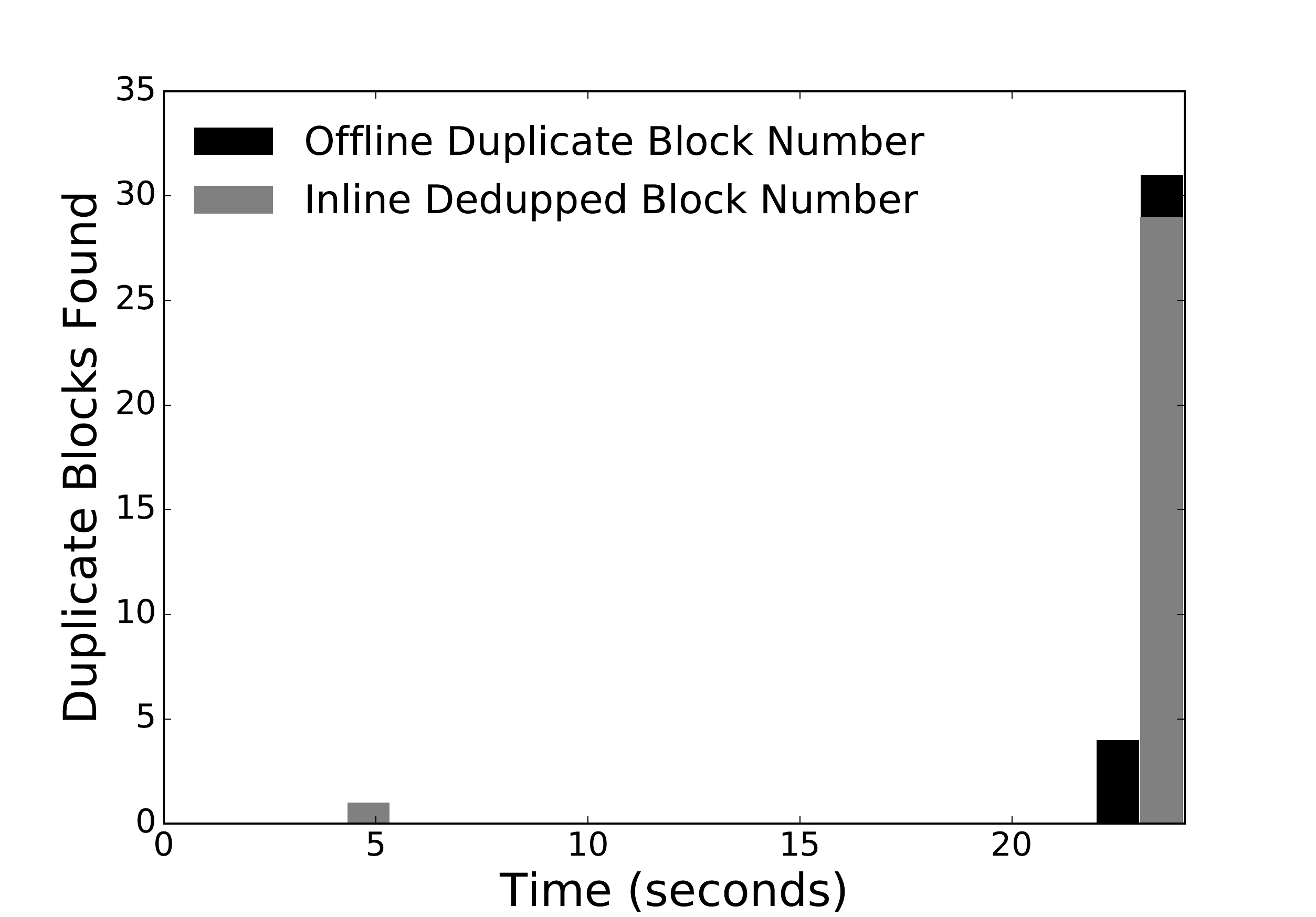}%
\label{linux_src}}
\caption{The duplicate blocks found in the VM image and Linux Kernel Source Code tar file. The x-axis is the time during writing the files while the y-axis is the number of duplicate blocks.}
\label{fig:inline_offline_H_type}
\vspace{-10pt}
\end{figure}

As shown in Table \ref{fcu}, HPDedup clearly outperforms the baseline and DIODE on the average hits of cached fingerprints. HPDedup outperforms DIODE by identifying data streams that have weak locality but do not belong to P-Type. It then avoids allocating cache space to these streams in order to make room for data streams with good locality.

This approach is effective. To show this, we compare the locality of the following two files in the Cloud-FTP workload: a Linux kernel 4.6 source code tar file and a VM image of CentOS 5.8 downloaded from OSBoxes. The two files are similar in size (2.7GB and 2.6GB). They are written to the primary storage after an inline deduplication process. The fingerprint cache size is set to 1\% of the data size (27MB and 26MB). Figure~\ref{fig:inline_offline_H_type} shows the number of duplicate blocks found in the two files.  Note that for the VM image, nearly all duplicate blocks can be found through the inline deduplication process.
DIODE treats both files as highly-deduplicatable (H-Type in DIODE). However, the number of duplicate blocks in the VM image file is significantly higher than that in the source code tar file. Moreover, DIODE ignores all P-Type files during inline deduplication by letting them to be processed during inline deduplication. However, multiple writes of the same P-Type files result in duplication and cannot be eliminated through differentiating file types. Different from DIODE, HPDedup allocates much less cache to the Cloud-FTP stream while writing the Linux Source Code tar file but allocates more when writing the VM image file. 

The result shows that simply using file type to guide cache allocation is insufficient. \textbf{HPDedup classifies data at finer-grained (stream temporal locality level) so that the efficiency of inline deduplication can be further improved.}


\subsection{Locality Estimation Accuracy}
HPDedup improves the efficiency of inline deduplication phase by allocating the fingerprint cache based on the temporal locality of each stream. Since \textit{LDSS} is an indicator which describes the temporal locality of data streams, it is critical to achieve accurate \textit{LDSS} estimation.

\begin{figure}[h]
\centering
\subfloat[\textbf{\emph{Normalized Observed LDSS}}]{\includegraphics[width=3.1in]{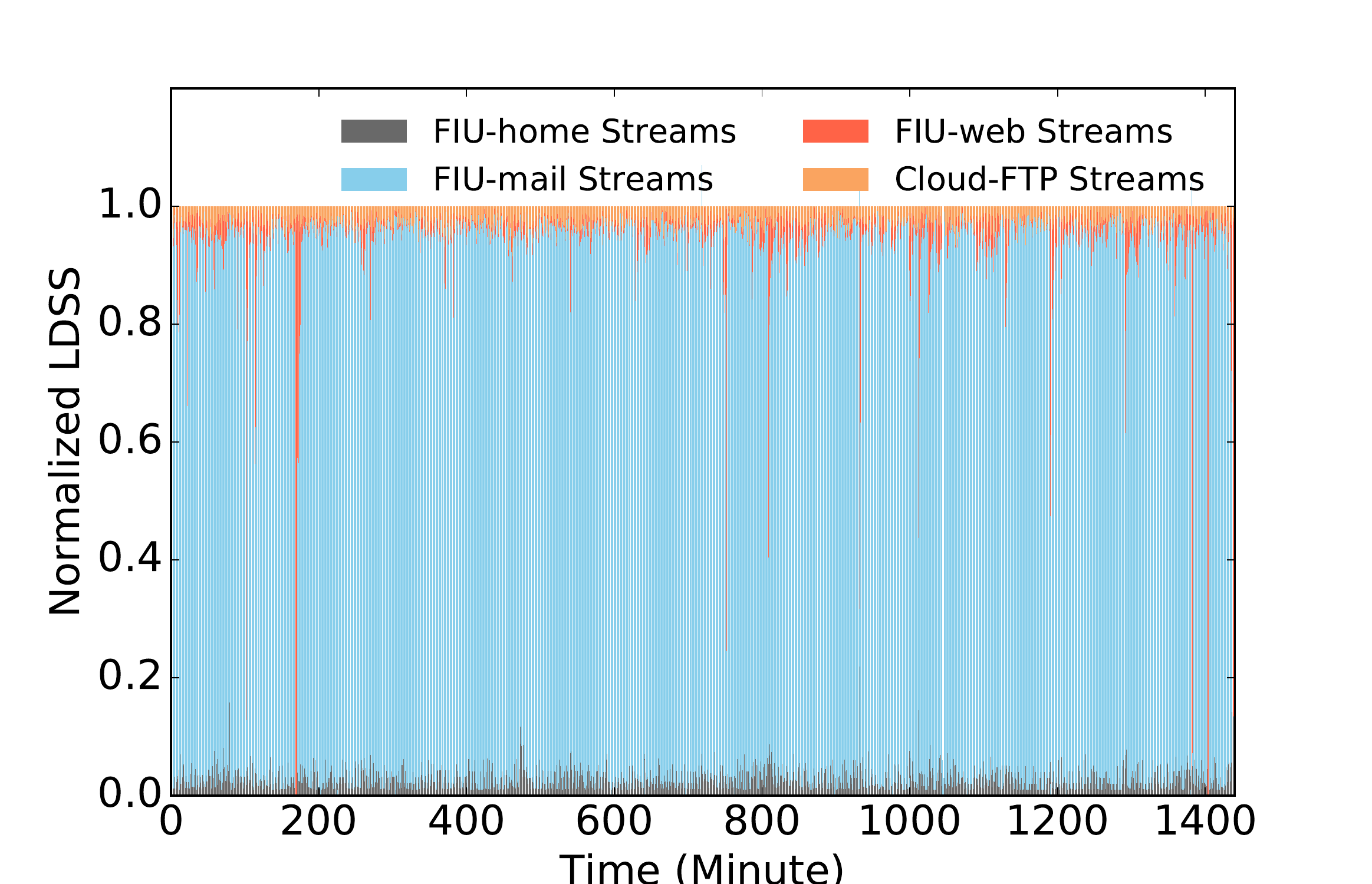}%
\label{1v1_observed_ldss}}
\vfill
\subfloat[\textbf{\emph{Cache Size Distribution (without LDSS estimation)}}]{\includegraphics[width=3.1in]{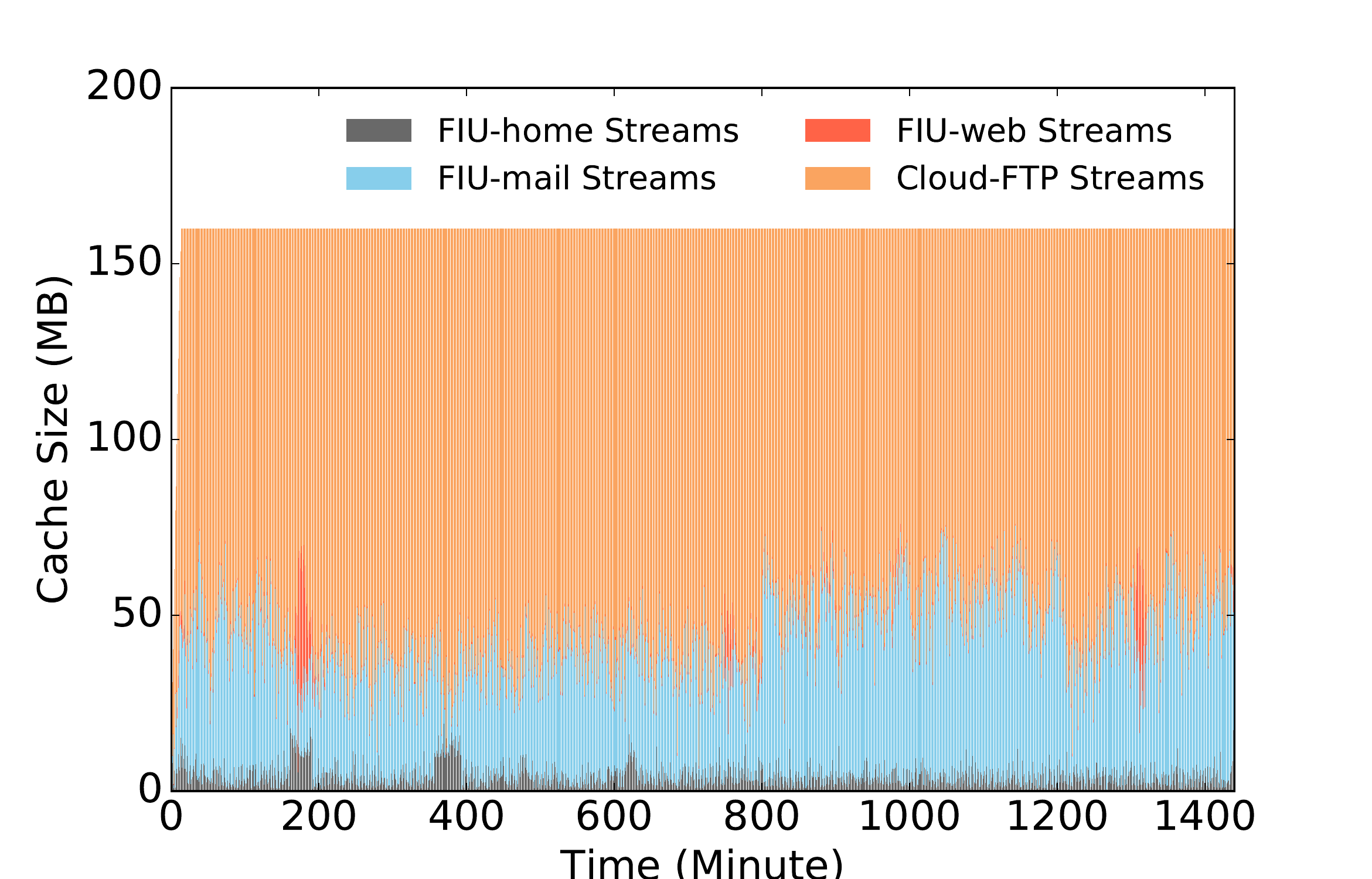}%
\label{1v1_idedup_cache_size}}
\vfill
\subfloat[\textbf{\emph{Cache Size Distribution (with LDSS estimation)}}]{\includegraphics[width=3.1in]{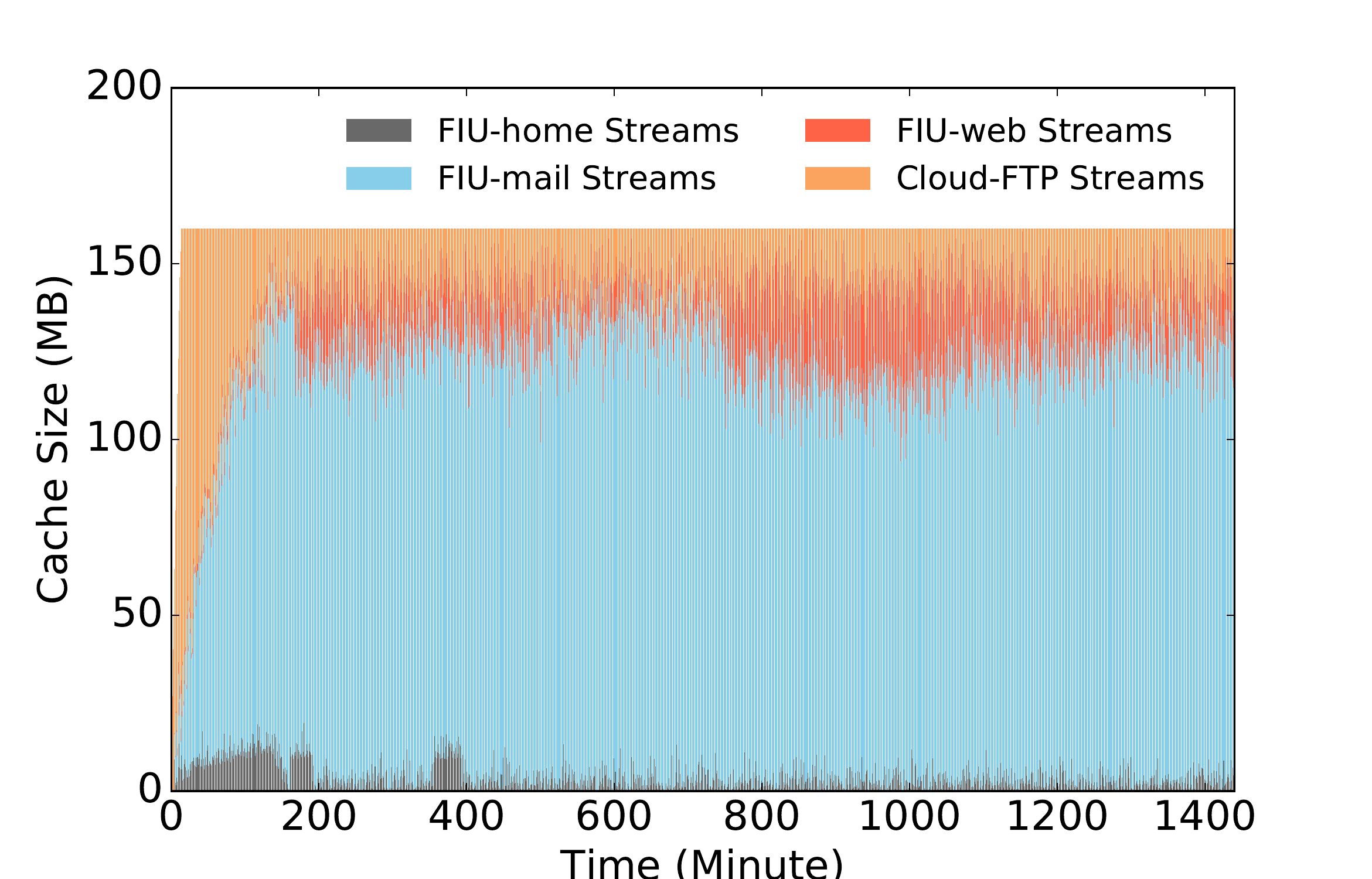}%
\label{1v1_hpdedup_cache_size}}
\caption{LDSS estimation accuracy.}
\label{estimation_accuracy}
\vspace{-10pt}
\end{figure}

Figure \ref{estimation_accuracy} shows the observed \textit{LDSS} for workload B. Here, the cache size is set to 160MB. To make the figures concise, the traces generated from the same template are aggregated. Figure \ref{1v1_observed_ldss} shows the observed \textit{LDSS} over time. The values of \textit{LDSS} are normalized. FIU-mail streams show the largest \textit{LDSS} thus indicating it has the best temporal locality. Nevertheless, as shown in Figure \ref{1v1_idedup_cache_size}, very little cache resource is occupied by FIU-mail streams when \textit{LDSS} estimation is not used. Cloud-FTP streams whose \textit{LDSS} is not high occupy the majority of cache resources. As shown in Figure \ref{1v1_hpdedup_cache_size}, with the guidance of \textit{LDSS} estimation, cache resource is cleverly allocated to streams based on their temporal locality. \textbf{LDSS estimation allocates fingerprint cache resources according to the temporal locality of streams and improves the inline deduplication ratio by 12.53\% (see Figure \ref{2v2}). The improvement clearly shows the effectiveness of \textit{LDSS} estimation in HPDedup.}

\subsection{Fragmentation}
\begin{figure}[t]
\vspace{-10pt}
\centering
\subfloat[\textbf{\emph{DIODE}}]{\includegraphics[width=0.42\textwidth]{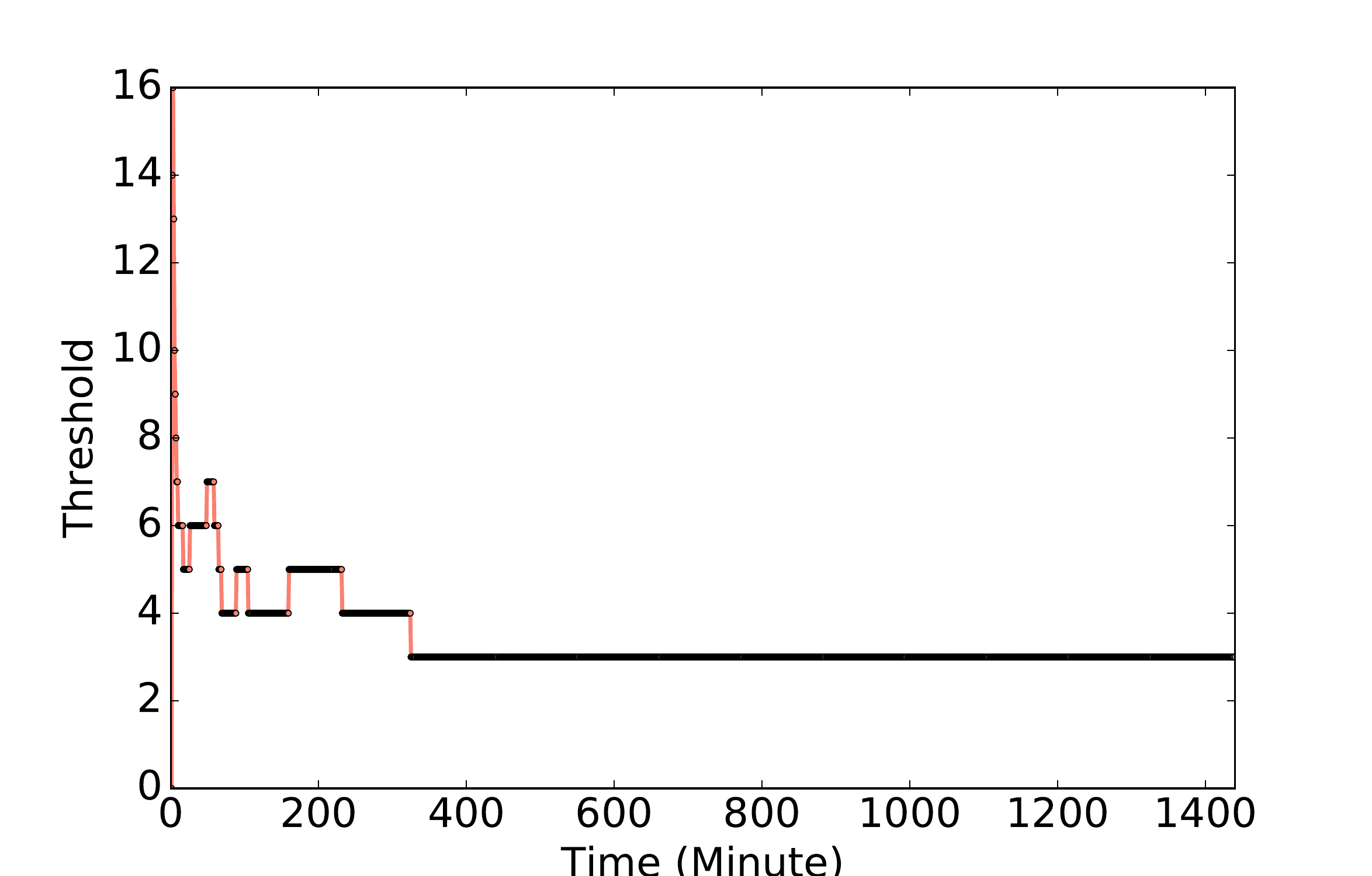}%
\label{diode_threshold}}
\vfill
\vspace{-10pt}
\subfloat[\textbf{\emph{HPDedup}}]{\includegraphics[width=0.42\textwidth]{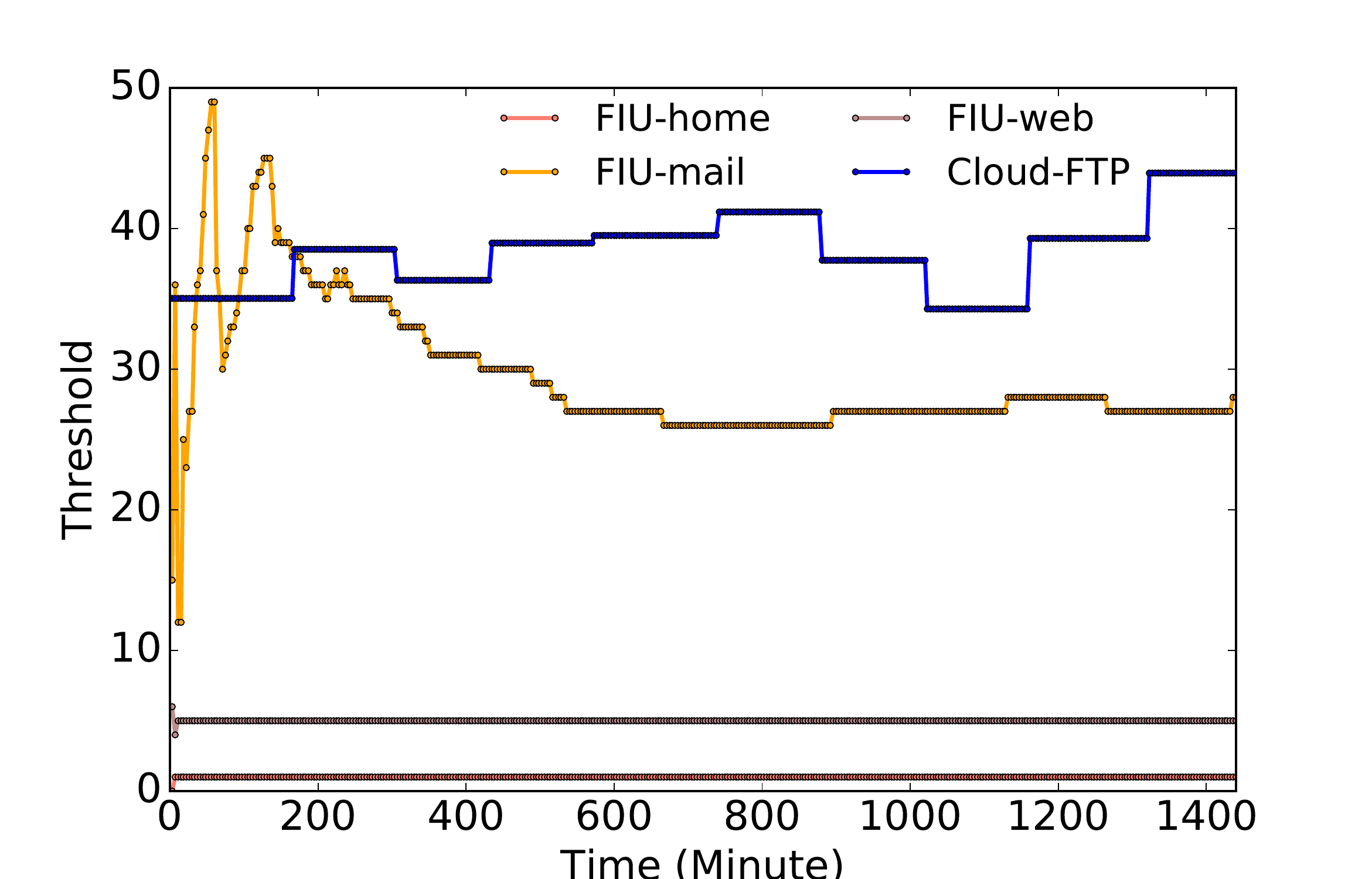}%
\label{hpdedup_threshold}}
\caption{Threshold vs. Time for HPDedup and DIODE (cache size: 160MB). }
\label{fig:fragmentation_profile}
\vspace{-10pt}
\end{figure}

In this subsection, we evaluate the fragmentation of files in storage system caused by deduplication. The length threshold of duplicate block sequence controls the fragmentation in both HPDedup and DIODE.

Both DIODE and HPDedup are able to adjust the threshold dynamically. Figure \ref{fig:fragmentation_profile} shows the threshold change along time for workload A in DIODE and HPDedup.

The inline deduplication ratio for DIODE and HPDedup are 57.62\% and 68.96\%, respectively. As one may see, HPDedup is able to adjust the threshold for each stream while DIODE uses a global threshold. The FIU-mail and Cloud-FTP have a higher threshold than FIU-home and FIU-web. Since larger threshold leads to less disk fragmentation, this result shows that \textbf{HPDedup introduces less fragmentation while achieving higher inline deduplication ratio than DIODE}.

\subsection{Overhead Analysis}

While HPDedup improves the efficiency of primary storage deduplication significantly, it inevitably incurs overhead. We analyze the overhead in this subsection. The overhead can be classified into computational overhead and memory overhead.

\subsubsection{Computational Overhead}
The computational overhead of HPDedup contains the following two parts: the histogram calculation time and the estimation algorithm execution time.

\begin{figure}[h]
\centering
\subfloat[\textbf{\emph{generating histogram for samples}}]{\includegraphics[width=1.68in]{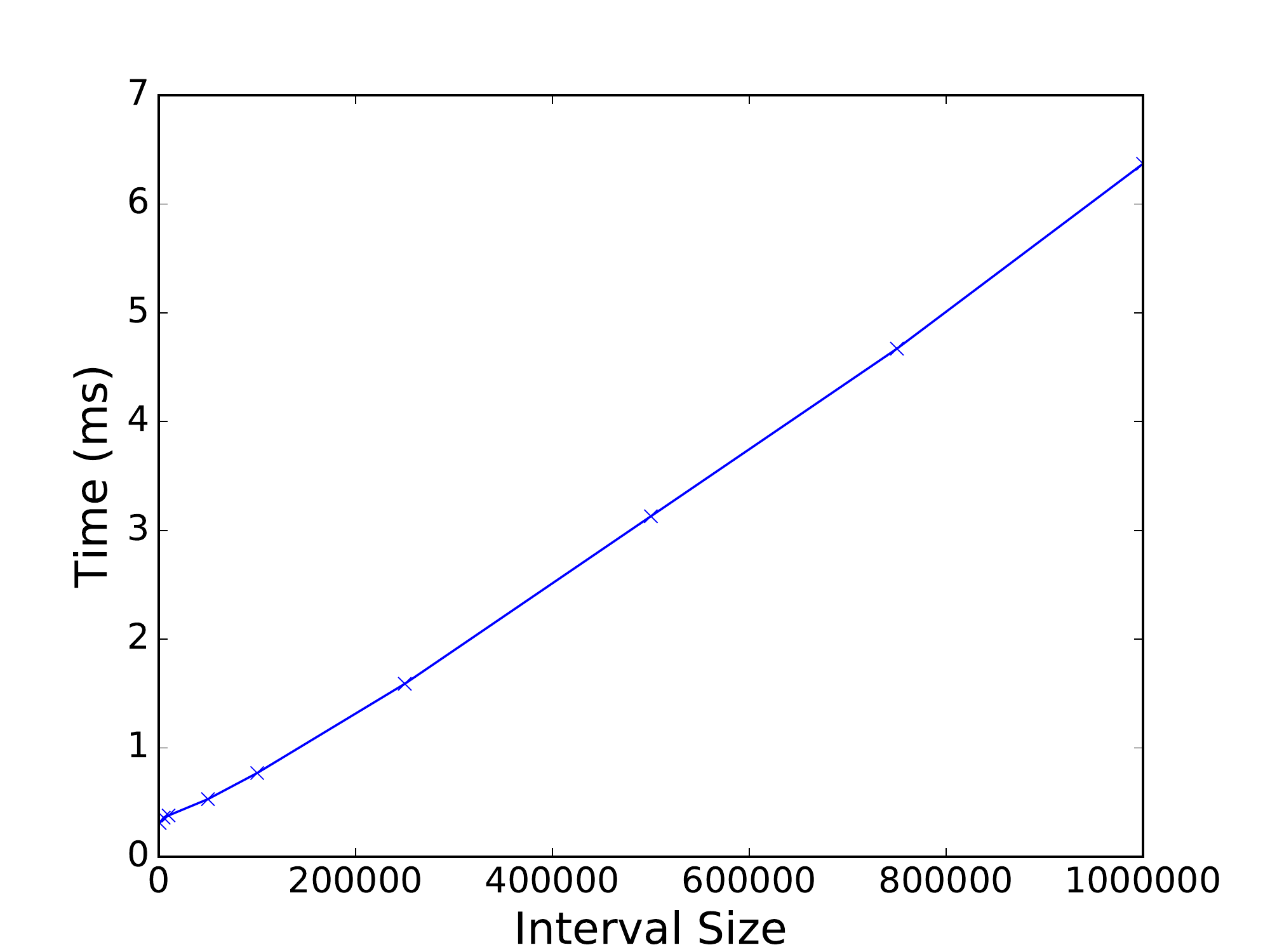}%
\label{cost_gen_tsf}}
\hfill
\subfloat[\textbf{\emph{temporal locality estimation}}]{\includegraphics[width=1.68in]{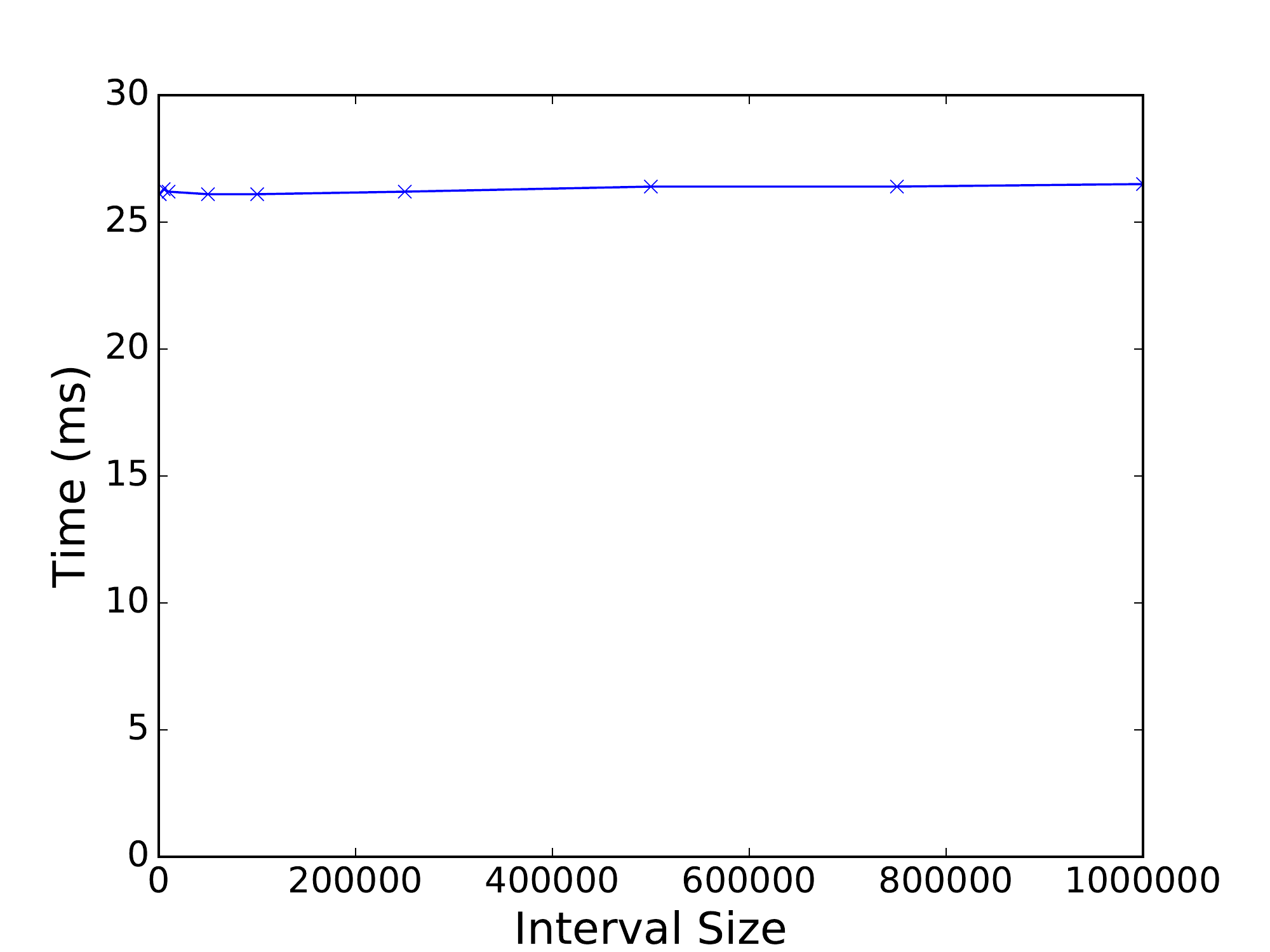}%
\label{cost_estimation}}
\caption{The time cost for HPDedup.}
\label{costs}
\end{figure}
To calculate the histogram, we only need to scan the sample buffer and add the count of the fingerprints to corresponding bins of the histogram. Therefore, the time complexity is $O(n)$ where $n$ is the number of samples. Figure ~\ref{cost_gen_tsf} shows the time used for generating the histogram in our current implementation. Here, the sampling rate is 15\%. We can see that the histogram calculation of an estimation interval with 1 million blocks takes less than 7ms.

Estimating the temporal locality of streams is achieved by using the method described in Section~\ref{subsec:temporal_locality}. The core of the estimation is to solve a linear programming problem. The linear programming problem can be solved in $O(n)$~\cite{megiddo1984linear} and even constant time~\cite{alon1990parallel} when the number of variables \emph{d} is fixed and the number of constraints is fixed. In our context, the condition is satisfied because too frequent duplicates in the sampling buffer will be used in a straightforward way during the estimation and will not be put into the linear programming. 

Note that the linear programming needs to be done for each stream. For every estimation interval, the temporal locality estimation takes about 26ms for each stream regardless the estimation interval size (see Figure \ref{cost_estimation}). The computing overhead is acceptable as the process is performed in background and does not affect the data write performance.

\subsubsection{Memory Overhead}
The primary memory overhead of HPDedup comes from the sampling buffer. With an estimation interval of size \emph{EI} and a sampling rate \emph{p}, the memory cost for tracking the histogram of samples is as below:
\[
 EI \cdot p \cdot (fpSize + counterSize)
\]
where $fpSize$ and $counterSize$ are the memory cost for storing fingerprints and the occurrence count, respectively. For instance, when the cache size is 160MB, there would be approximately 2.62M cache entries. For the sampling rate of 15\%, the memory overhead is only 4.49MB (2.81\% of cache size) even if we choose a large estimation interval factor (e.g., Workload C, 0.6). In practice, for the mixed streams with better temporal locality, the memory overhead is much less (e.g., 2.25MB for Workload A and 2.99MB for Workload B) as the \textit{estimation interval factor} can be set to smaller values. \textbf{Compared with the improvement (19.80\% - 25.81\%) of inline deduplication ratio for the three workloads with 160MB cache size),  the memory overhead of HPDedup is acceptable.}

\section{Related Work}\label{sec:related}
\subsection{Primary Storage Deduplication Mechanisms}
Data deduplication achieves a great success in backup storage systems. Recent research exploit various ways to apply deduplication in primary storage systems for both reducing data size in storage devices and improving I/O performance. Existing work can be classified into three categories: \textit{Inline primary storage deduplication}, \textit{Post-processing/Offline primary storage deduplication} and \textit{Hybrid inline and post-processing deduplication}.

\textbf{Inline primary storage deduplication.} Most inline primary deduplication exploits the locality in primary workloads to perform non-exact deduplication. iDedup~\cite{srinivasan2012idedup} exploits the temporal locality by only maintaining an in-memory cache to store the fingerprints of data blocks. To exploit the spatial locality, iDedup only eliminates duplicates in long sequences of data blocks. POD \cite{mao2014pod} aims at improving the I/O performance in primary storage systems and mainly performs deduplication on small I/O requests. HANDS~\cite{wildani2013hands} uses working set prediction to improve the locality of fingerprints. Koller et al.~\cite{koller2010deduplication} uses content-aware cache to improve the efficiency of I/O by avoiding the influence of duplicated data. PDFS \cite{yu2016pdfs} argues that the locality may not commonly exist in primary workloads. To avoid the disk bottleneck of storing fingerprint table, a similarity based partial lookup solution is proposed. Leach \cite{lin2014leach} exploits the temporal locality of workloads by a splay tree. These work do not consider scenarios involving VMs and containers in the cloud where workloads for the primary storage contain a mix of data streams with different access patterns.

\textbf{Post-processing/Offline primary storage deduplication}. Post-processing deduplication performs deduplication during the idle time of primary storage systems. Ahmed El-Shimi et al. \cite{el2012primary} propose a post-processing deduplication method built in Windows Server operating systems. Similar with HPDedup, DEDIS \cite{paulo2016efficient} is built in the Xen hypervisor to provide data deduplication functionality to multiple virtual machines. The main purpose of post-processing primary storage deduplication is to avoid the high I/O latency introduced by inline on-disk dedupe metadata lookup. However, even though the locality does not always exist in primary workloads, it is much more efficient to use inline caching rather than post-processing to eliminate duplicates in the portion of workloads with decent locality. The contribution of HPDedup is to differentiate the deduplication procedure for primary workloads according to the temporal locality of workloads.

\textbf{Hybrid inline and post-processing deduplication}. Combining inline and post-processing deduplication together has been exploited by RevDedup  \cite{li2015efficient} in backup storage deduplication to improve the space efficiency. For primary storage deduplication, DIODE \cite{tang2016diode} also proposes a dynamic architecture of inline-offline deduplication. Like ALG-Dedupe \cite{fu2014application}, DIODE is an application-aware deduplication mechanism. File extensions are classified into three types according to their potential deduplication ratio. Moreover, whether performing inline deduplication on a file is determined by the extension of the files. However, our experiments show that file types are not sufficient for achieving good inline deduplication performance and there is a lot of room to improve. The key difference between HPDedup and DIODE is that HPDedup gives a dynamic locality estimation method to improve inline deduplication performance, therefore reduces the load of the more expensive post-processing deduplication process.
\subsection{Unseen Distribution Estimation}
Estimating the number of duplicates in a time frame for a data stream is similar to estimating the distinct elements in a large set, for which various statistics based methods (e.g.,~\cite{haas1995sampling,mao2007estimating}) have been investigated. Fisher et al.~\cite{fisher1943relation} describe a method to estimate the number of unknown species given a histogram of randomly sampled species. Theoretical computer science community has been trying to address how to perform the estimation with less samples~\cite{bar2001sampling,batu2005complexity,valiant2011testing,guha2006streaming}. Recently, this line of work has been extended by Valiant and Valiant~\cite{valiant2011estimating} to characterize unobserved distributions. They prove that only $O(\frac{n}{log(n)})$ samples are sufficient to provide an accurate estimation of the whole dataset, in which $n$ is the size of the whole dataset. Harnik et al.~\cite{harnik2016estimating} utilizes the theory to estimate the duplicates in storage systems.
\subsection{Dynamic Flash Cache Management}
Using prediction or historical information of workloads to improve the cache efficiency has been explored in flash cache management \cite{arteaga2016cloudcache,yang2013hec,huang2016improving,liu2014plc,koller2015centaur}. These work studied cache admission policies and dynamic cache allocation to reduce the flash wear-out. To the best our knowledge, HPDedup is the first work to use locality estimation to deal with the cache contention problem in inline deduplication for primary storage.

\section{Conclusion}\label{sec:concl}
In scenarios where multiple virtual machines or containers running in the cloud,  many applications are placed in the same physical machine. Removing duplicate I/Os from the primary storage in these scenarios is useful to both improve the capacity efficiency and I/O performance. We proposed HPDedup, a hybrid prioritized deduplication method for primary storage in the cloud. HPDedup used a dynamic temporal locality estimation algorithm to achieve high inline cache efficiency and left the relatively small number of duplicates that were not in the cache to the post-processing deduplication phase to handle. By doing so, HPDedup was able to achieve exact deduplication in primary storage systems.  
Comparing to the state-of-art inline deduplication methods, HPDedup significantly improved the inline cache efficiency therefore achieved high inline deduplication ratio. HPDedup improves the inline deduplication ratio by up to 39.70\% compared with iDedup in our experiments. Meanwhile, the improved cache efficiency made the post-processing deduplication process less a burden for the performance of an inline primary deduplication system.  For example, HPDedup reduces up to 45.08\% disk capacity requirement compared with the state-of-art post-processing deduplication mechanism in our evaluation.

\section*{Acknowledgment}
We would like to thank Gregory Valiant from Stanford University for helping us to further understand the unseen entropy estimation algorithm. This work is partially supported by the The National Key Research and Development Program of China (2016YFB0200401), by program for New Century Excellent Talents in University， by National Science Foundation (NSF) China 61402492, 61402486, 61379146,  by the laboratory pre-research fund (9140C810106150C81001).

{
\scriptsize
\bibliographystyle{IEEEtran}
\newcommand{\BIBdecl}{\setlength{\itemsep}{0.25em}}
\bibliography{stagefs}
}

\end{document}